\documentclass[aps,prl,onecolumn,superscriptaddress,notitlepage]{revtex4-1}

\usepackage{graphicx,color}% Include figure files
\usepackage{epsfig}
\usepackage{dcolumn}% Align table columns on decimal point
\usepackage{amsmath,amssymb,bm}

\usepackage{hyperref,url}
\usepackage[noabbrev]{cleveref}
\usepackage{CJK}
\usepackage{lineno}
\usepackage[english]{babel}
\usepackage[T1]{fontenc}
\usepackage[normalem]{ulem}
\usepackage{pifont}
\usepackage{amsfonts}
\usepackage{wasysym}
\usepackage{amsbsy}
\usepackage{psfrag}
\usepackage{color}
\usepackage{multirow}
\usepackage{cases}
\usepackage{stackengine}
\usepackage{float}
\usepackage{blkarray}
\usepackage{cancel}
\usepackage{mathrsfs}
\usepackage{empheq}
\usepackage{lipsum}

\usepackage{titlesec}
\titlespacing*{\section}{0pt}{0.2\baselineskip}{0\baselineskip}

\usepackage{xr}

\usepackage{gensymb}
\usepackage{newtxtext}
\usepackage{newtxmath}
\usepackage[dvipsnames]{xcolor}

\addto\captionsenglish{}

\DeclareMathAlphabet{\textbfsf}{\encodingdefault}{\sfdefault}{bx}{sl}

\hypersetup{
    colorlinks=true,
    linkcolor=blue,
    citecolor=blue,
    urlcolor=blue,
}

\usepackage{fancyhdr}
\begin{document}

\setlength{\belowdisplayskip}{0pt} \setlength{\belowdisplayshortskip}{0pt}
\setlength{\abovedisplayskip}{0pt} \setlength{\abovedisplayshortskip}{0pt}

% \linenumbers

%Title of paper
\title{Arrested coalescence drives helical coiling and networking of filamentous smectic condensates}

%\author{List of authors}
\author{Christopher A. Browne}
\email{cabrowne@umich.edu}
\affiliation{Department of Chemical Engineering, University of Michigan, Ann Arbor, MI 48109, USA}
\affiliation{Department of Chemical and Biomolecular Engineering, University of Pennsylvania, Philadelphia, PA 19104, USA}

\author{Paul G. Severino}
\affiliation{Department of Physics and Astronomy, University of Pennsylvania, Philadelphia, PA 19104, USA}

\author{Yvonne Zagzag}
\affiliation{Department of Chemical and Biomolecular Engineering, University of Pennsylvania, Philadelphia, PA 19104, USA}
\affiliation{Department of Physics and Astronomy, University of Pennsylvania, Philadelphia, PA 19104, USA}

\author{Jacob Z. Cloutier}
\affiliation{Department of Mechanical Engineering, University of British Columbia, Vancouver, BC V6T 1Z2, Canada}

\author{Aaron C. Boyd}
\affiliation{Department of Chemical and Biomolecular Engineering, University of Pennsylvania, Philadelphia, PA 19104, USA}

\author{Yihao Chen}
\affiliation{Department of Physics and Astronomy, University of Pennsylvania, Philadelphia, PA 19104, USA}

\author{Arjun G. Yodh}
\affiliation{Department of Physics and Astronomy, University of Pennsylvania, Philadelphia, PA 19104, USA}

\author{Chinedum O. Osuji}
\email{cosuji@engineering.upenn.edu}
\affiliation{Department of Chemical and Biomolecular Engineering, University of Pennsylvania, Philadelphia, PA 19104, USA}
%\email{cosuji@seas.upenn.edu}

\begin{abstract}

Liquid-liquid crystal phase separation (LLCPS) occurs in many industrial and biological settings. Interestingly, when smectic phases demix from the homogeneous mixture, they can form filamentous condensates that spontaneously assemble into sparse networks with rich life-like dynamics. Here, we study the underlying process of filament networking. Microscopy reveals that new linkages between filaments are initiated by an adhesive interaction between straight filaments; parallel filaments snap into contact and then rapidly wind into helical coils, despite the absence of molecular chirality or transitions between mesophases. Using polarized optical microscopy, theoretical modeling, and simulation, we show that parallel filaments coalesce, but are arrested in double-barreled ``ribbon'' structures. This arrested coalescence is driven purely by interfacial energy minimization under the constraints of smectic layering. The arrested ribbons spontaneously coil into double helices to further reduce interfacial area, thus driving compaction into networks. We propose a microstructure consistent with this interpretation, which quantitatively predicts the extent of arrested coalescence. In total, these findings suggest a generic pathway for network formation in liquid crystals that provides insight about the formation of condensate networks in other engineered or biological materials.
\end{abstract}

\date{\today}

\maketitle

\subsection*{Introduction}

Liquid-liquid phase separation (LLPS) plays key roles in the processing of engineered materials \cite{liu2013polymerization,cui2022dynamic} , formation of condensates in living cells \cite{shin2017liquid,fraccia2021liquid,julicher2024droplet}, and the production of biomaterials by living organisms \cite{harrington2024fluid,almohammadi2025liquid}. Some of these systems can form condensates with liquid crystalline (LC) order, which generally exhibit phase separation dynamics and condensate morphologies that differ from that in isotropic fluids \cite{fraccia2021liquid,harrington2024fluid,almohammadi2025liquid}. The elasticity and defects of liquid crystals can alter the morphology of demixed droplets, giving rise to tactoids \cite{alsayed2004melting,bagnani2018amyloid,scheff2020tuning,fraccia2021liquid, almohammadi2025liquid}, ribbons\cite{dogic2006ordered}, tori \cite{tortora2010self}, membranes \cite{adkins2025topology}, and surface facets \cite{jeong2014chiral} in systems with nematic, cholesteric, or columnar ordering. Additionally, the dynamics of the phase separation process can be slowed or halted by the elasticity energetics and defect structure \cite{patel2021temperature,patel2023long,lagerwall2024good}. Lastly, the morphology of these metastable condensate drops are strongly influenced by thermal and processing history \cite{almohammadi2023evaporation,almohammadi2023disentangling} and hydrodynamics \cite{almohammadi2020flow,almohammadi2022shape} during condensate nucleation and growth. Examples of kinetically-trapped architectures that arise in this context include composite gels with entrained colloids \cite{anderson2001cellular} or isotropic droplets \cite{reyes2019isotropic}, liquid crystal shells around liquid droplets \cite{vitelli2006nematic,liang2011nematic,kim2018self,noh2021topological}, and core-shell fibers \cite{buyuktanir2010self,wang2016morphology,sharma2018electrospun}. 

This contribution focuses on liquid-liquid crystal phase separation (LLCPS) wherein smectic liquid crystals form upon demixing. The resulting rich phenomenology is driven in large part by the constant layer spacing of smectic LCs \cite{Lacaze2025PRL} that imposes constraints on growth. Some of these systems form highly-elongated metastable filaments \cite{naito1995preferred,naito1997pattern,todorokihara2004periodic}, wherein elongation at constant radius $R_\text{F}$ \cite{naito1997pattern} is favored over sustained growth of spherical nuclei. This behavior arises above a threshold size when strong homeotropic anchoring penalizes the nucleation of new smectic sheathing layers \cite{pratibha1992cylindrical,arora1989reentrant,weinan1999dynamics}. These filaments are reminiscent of other stable and metastable filamentous structures observed in other nematic \cite{wei2019molecular,peddireddy2021self}, re-entrant smectic \cite{weinan1999dynamics,arora1989reentrant,pratibha1992cylindrical}, and lyotropic smectic \cite{zou2006stability,huang2006theory,reissig2010three} systems, though the underlying driving forces for pattern formation in these systems are distinct. 

Filamentous smectic condensates can exhibit a rich phenomenology of non-equilibrium dynamics, driven in large part by the metastability of droplet morphologies \cite{morimitsu2024spontaneous,browne2025diversity}. Rapidly growing filaments generate active hydrodynamic flows, leading to cyclic buckling and densification of filaments \cite{shelley2000stokesian}. Under this hydrodynamic confinement, filaments reconfigure when sufficiently dense into lower energy condensate morphologies, including flat drops\cite{morimitsu2024spontaneous}, coils \cite{browne2025diversity}, and  tori \cite{naito1995preferred}. In some cases, these rapid reconfiguration events can instead give rise to the formation of sparse networked architectures, which can sustain active dynamics for hours \cite{morimitsu2024spontaneous}. It remains unknown what sets the sparsity or topology of these condensate networks, which can vary with mesogen composition, solvent composition, thermal quench rate, and confinement \cite{morimitsu2024spontaneous,browne2025diversity}, in large part because the reconfiguration of dense filaments into network linkages remains poorly understood.

In this work, we investigate the route to networking in filamentous smectic condensates. Surprisingly, we observe that network formation is mediated by spontaneous coiling of initially-disconnected parallel filaments into double-helical coils, which then form the nascent network linkages. Formation of these coils is not driven by any underlying molecular characteristics, such as chirality or bend \cite{OlegJakliSelingerBentCore,jakli2000helical}; coil formation arises at all temperatures within the binodal. Instead, high speed imaging reveals that coiling is initiated by the rapid snapping-together of co-aligned filaments into transient ``ribbon'' morphologies. Full-coalescence of filaments is arrested in these ribbons, presumably by elastic constraints on the liquid crystal microstructure. Once formed, the ribbons wind with a spontaneously-selected macroscopic chirality. Geometric analysis of these coils reveals that they effectively reduce interfacial energy; reminiscent of entropically-driven coiling of filaments with low torsional and bend rigidity \cite{snir2005entropically}. We propose a set of internal smectic microstructures that can rationalize the stabilization of these arrested droplet morphologies. By isolating single filaments within microdroplets and tracking their shape transformations, we show that coiling is driven by mesophase structure rather than emergent activity of network growth. Our results highlight how condensates can adopt complex morpholgies driven purely by interfacial stresses under the constraints of liquid crystal ordering. These phenomena could play unexplored roles in the morphological transformations or networking of condensates in other engineered and biological materials \cite{pombo2024membrane,hnisz2017phase,shen2021effects,basu2025vimentin}.

% -----------------------------------
\section*{Results}

\begin{figure}
    \centering
    \includegraphics[width=\linewidth]{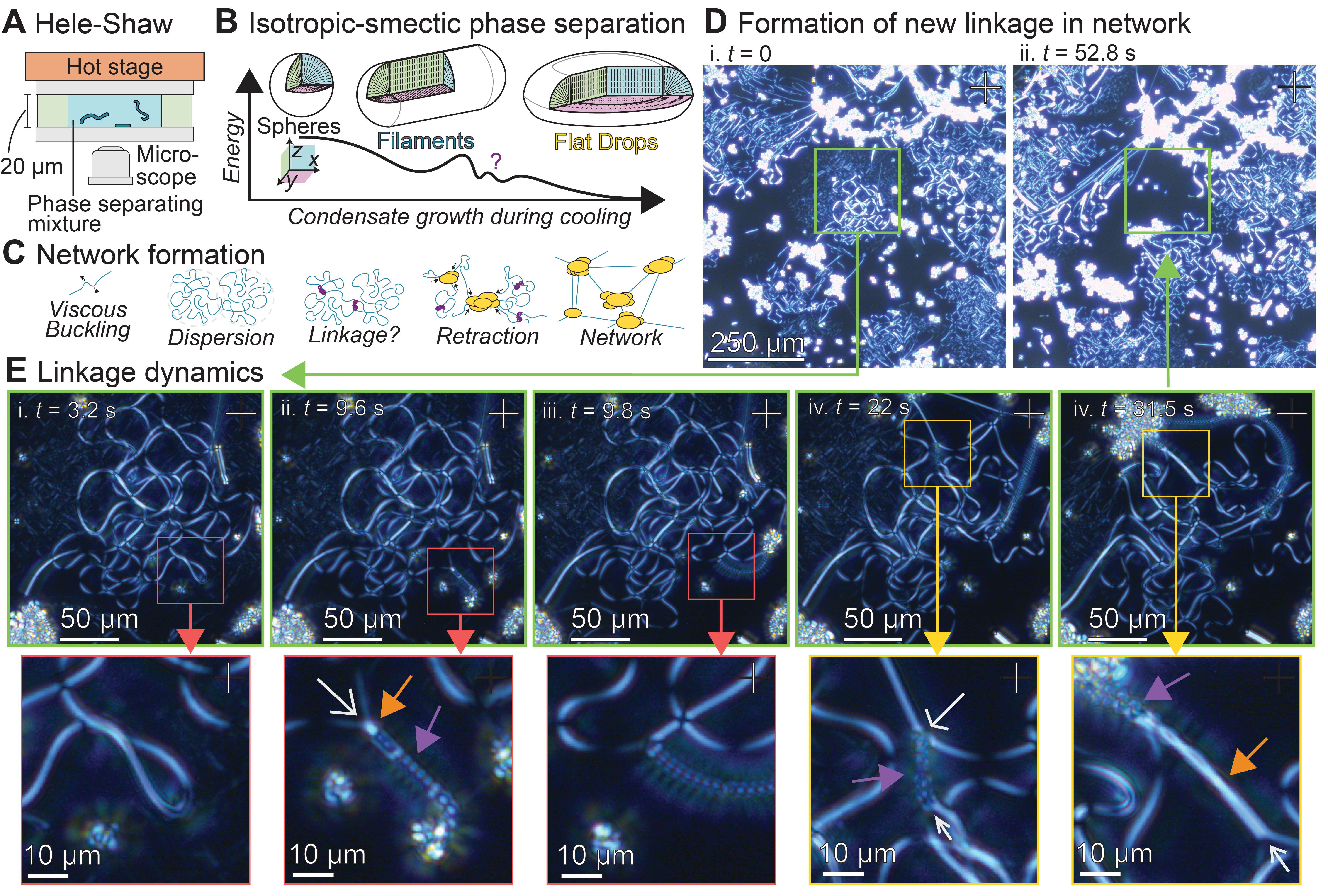}
    \caption{\textbf{Filament coiling mediates networking in Hele-Shaw cells.} \textbf{A} 12OCB and Squalane mixture within a 20 $\muup$m gap Hele-Shaw cell is cooled into the binodal to induce demixing  as previously described \cite{morimitsu2024spontaneous}. \textbf{B} Schematic energy landscape of metastable filament transformations to form unknown intermediate structures before forming flat drops. \textbf{C} Schematic of network formation, which occurs when filaments above threshold density collpase and form new linkages, which coarsen to aggregate nodes composed of flat drops. \textbf{D} Imaging on CMOS (ThorLabs) under
crossed polarizers shows birefringence of smectic condensate early in network formation at 5$\times$ magnification before (i) and after (ii) a collapse event. \textbf{E} Zoomed view of collapse event (20$\times$ magnification) shows nascent linkage forming between initially disconnected filamentous masses. Lower panel shows zoomed insets in red and yellow. Free filaments meet at Y-junctions (white arrows) forming bright contact regions that zipper filaments together (orange arrows) and subsequently form an interdigitated, ribbed structure (purple arrows).}
    \label{fig:1}
\end{figure}

\noindent\textbf{Phase separating filaments link to form spatial networks in Hele-Shaw cells.} We prepare binary mixtures of a smectogen (12OCB) and solvent (squalane, "SqA") at $\omega=0.3$ to $0.45$ mass fraction smectogen at elevated temperatures $T>80\degree$C, where the components are fully miscible and form an isotropic liquid at all compositions, and load the heated liquid into $20~\muup$m gap glass Hele-Shaw cells (Methods). We equilibrate imaging cells at $T = 80\degree$C before slowly cooling into the binodal region at a rate of $0.1$ to $0.5\mathrm{\degree C/min}$ using a temperature control imaging stage (\textit{Linkam Scientific} THMS600). We image the phase separation using an inverted microscope under bright field, polarized, and phase-contrast imaging (Methods, Fig. \ref{fig:1}A).

Upon phase separation, we observe the condensation of dispersed smectogen-rich droplets with smectic A liquid crystalline ordering dispersed within a solvent-rich isotropic liquid as the continuous phase. Consistent with previous reports, these smectic condensates grow as filaments that elongate at constant radius  \cite{naito1995preferred,naito1997pattern,todorokihara2004periodic}. Under hydrodynamic confinement, metastable filaments overcome energy barriers to rearrange into lower energy morphologies, including flat drops (Fig. \ref{fig:1}B), which network into a sparse 2D network (Fig. \ref{fig:1}C) \cite{morimitsu2024spontaneous,browne2025diversity}. It remains unknown how these rapid reconfiguration events give rise to network linkages---and consequently what determines the network architecture.

We image the dynamics of one such representative linkage process at varying magnification under crossed polarizers (Movie 1). At an early time in network formation, dense filaments interpenetrate as they densify (green region in fig. \ref{fig:1}Di). Higher magnification of this region shows the formation of a new node within this dense mass, resulting in new linkages between the previously disconnected network nodes (Fig.  \ref{fig:1}E).  One portion of bent filament (Fig. \ref{fig:1}Ei red inset) suddenly snaps into contact with itself to form a ribbed, interdigitated structure (Fig. \ref{fig:1}Eii purple arrow). At the leftmost edge of the ribbed structure is a bright region where the filament self-contacts in a Y-junction with the two free ends of the filament (orange arrow). The bright self-contacting region and the ribbed region both propagate, drawing more of the free filament into the ribbed structure (Fig. \ref{fig:1}Eiii purple and orange arrows). A subsequent collapse event nearby (yellow inset, panels iv--v) shows a similar ribbed structure forming an X-linkage between two co-aligned portions of filaments (purple arrow panel iv). Again, a bright contact region propagates ahead of the ribbed region (orange arrow panel v). A cascade of similar events results in a dense aggregate mass, which forms a new node within the condensate network that provides new node linkages (Fig. \ref{fig:1}Dii). Initiation of new linkages throughout the network are predominantly driven by similar cascade events. We thus study these collapse events using high speed imaging to discern the critical conditions leading up to collapse and the subsequent cascade of morphological rearrangements.

\begin{figure}
    \centering
    \includegraphics[width=6.5in]{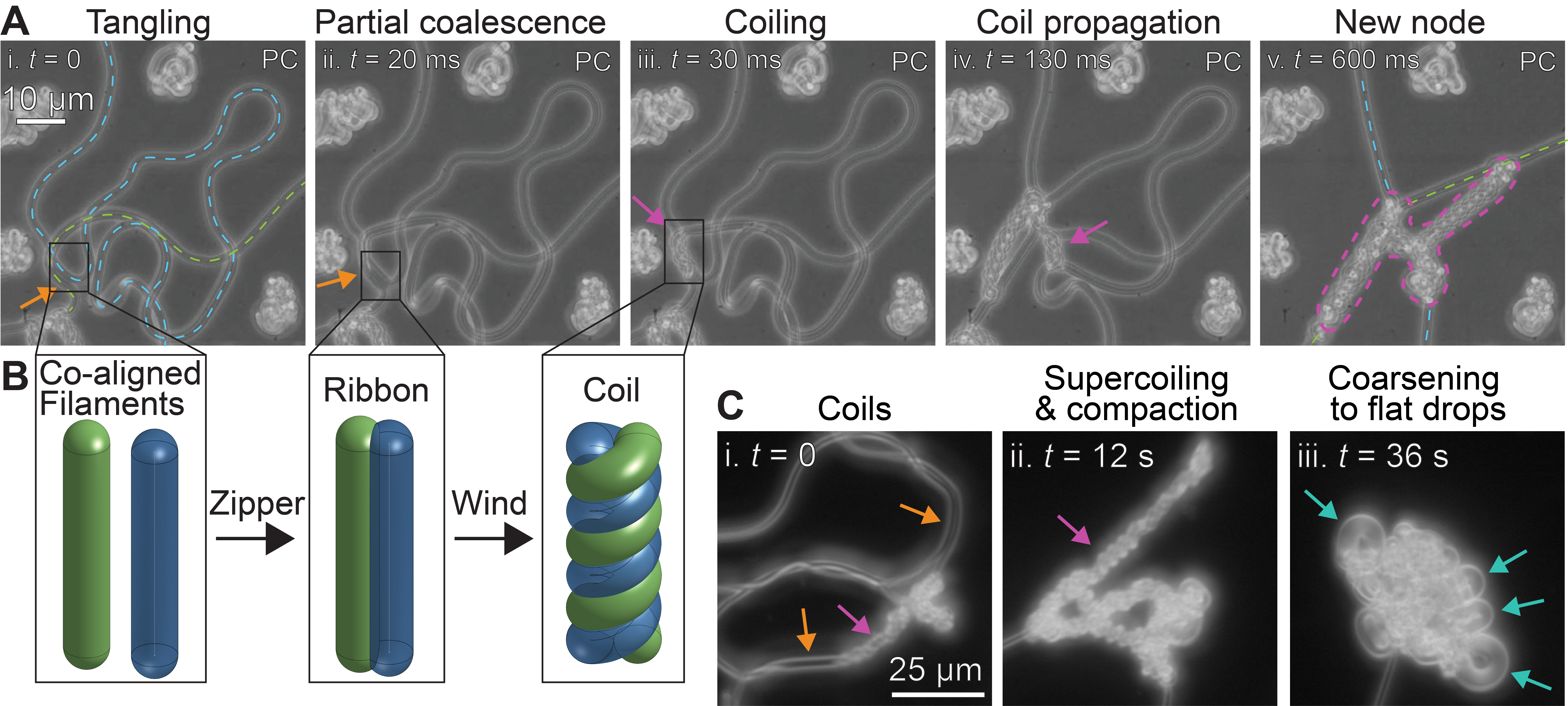}
    \caption{\textbf{Filaments link by partially coalescing and winding into helical coils.} \textbf{A} Representative linkage process captured by high speed camera (1000 fps; \textit{Phantom} VEO 1010) under 63$\times$ magnification. Phase contrast imaging shows the scattering interface of the condensate as bright. \textbf{B} Rendering of two initially-distinct filaments, colored green and blue to visually distinguish, zippering to form ribbon structure and winding to form coil structure. \textbf{C} Phase contrast imaging of coiled regions compacting into a supercoiled aggregate and coarsening into flat drops.}
    \label{fig:2}
\end{figure}

\noindent\textbf{Linkages mediated by arrested coalescence and helical coiling of filaments}. We record linkage events using high speed phase contrast imaging (1000 fps; \textit{Phantom} VEO 1010) at 63$\times$ magnification. Movie 2 shows one representative process whereby two initially-disconnected filaments (blue and green dashed contours in fig. \ref{fig:2}A panel i) link to form a new connected aggregate node (purple outline in panel v). As the two initially-disconnected filaments grow, one section co-aligns and makes incipient contact (orange arrow panel i). The two filaments rapidly snap together (orange arrow panel ii), forming a partially-coalesced structure with a double-barreled ``ribbon'' structure (schematic in fig. \ref{fig:2}B). This partially-coalesced region propagates outwards, zippering the filaments together with an adhesive interaction that increases the filament-filament contact area.

As this zippering process continues, the ribbon structure suddenly winds, resulting in a locally-coiled region that traces out a double helix (purple arrows in fig. \ref{fig:2}A panels iii and iv; rendering fig. \ref{fig:2}B). Once initiated, the coil also propagates rapidly, drawing in the ribbon of partially-coalesced filaments to form a dense double-helical coil. The rapid retraction of filaments into the denser coil initiates a cascade of additional adhesive contacts, which wind to form additional coiled regions (Fig. \ref{fig:2}A iv). We observe coils with both handedness, with no preference in chirality. 

Some coils persist in this state for long periods, often exceeding minutes. Others, including the coil pictured here, continue to twist and compact, supercoiling to form a dense aggregate node (end of Movie 2; Fig. \ref{fig:2}A panel v). We observe these aggregates coarsening into the previously-reported flat drop structures characteristic of the macroscale network nodes (blue arrows in fig. \ref{fig:2} C), though the disordered 3D nature of the supercoiled aggregates precludes direct characterization of the coarsening pathway. Throughout the network, we observe that new nodes are predominantly formed by similar rapid sequences of ribbon zippering and helix coiling, highlighting the importance of this process in forming interconnected condensate networks. We furthermore observe a wide range of zippering and coiling events that occur independent of network formation---including the formation of free-floating helical coils (Fig.\ref{SIfig:freefloat}, Movies 3 and 4)---indicating that arrested coalescence and helical coiling is a generically spontaneous process. 

The origin of this spontaneous ribbon zippering and coil winding is not immediately apparent. The formation of helical structures in liquid crystalline materials is often associated with molecular chirality; or molecules with a rigid bent core (sometimes termed ``banana'' molecules), which can exhibit exotic mesophases (e.g. B7) with spontaneous symmetry breaking that can result in chiral structures \cite{reddy2002helical, coleman2003polarization,reddy2006bent,jakli2000helical}. Our smectogen 12OCB is neither chiral nor strongly bent, so we do not expect such mesophases to arise. Furthermore, these chiral mesophases are not preceded by filament zippering into ribbons, as we observe here. Nevertheless, we can be certain that coils are not formed due to a transition to a lower-energy chiral or B7 mesophase because of the observed sequence of morphology transformations. We widely observe chiral coils transform into flat drops (Fig. \ref{fig:2}C), whose mesophase structure has been characterized and is known to be the original achiral smectic A \cite{morimitsu2024spontaneous}; thus, the intermediate coil conformation must also be in the smectic A phase. 

Consistent with this interpretation, we see co-existence of coils with filaments at all temperatures within the binodal. We do not observe any critical temperature beneath which coils become increasingly favored, nor do we ever observe filaments spontantously coil in the absence of physical or hydrodynamic confinement, as might be expected for achiral mesophase transition. Instead, we exclusively see coils form when co-aligned filaments grow sufficiently close by densification to ``snap'' into contact, after which chirality is spontaneously selected with no preference in handedness. These observations suggest that coiling is purely due to interfacial end elastic interactions between smectic A condensate filaments.

\noindent\textbf{Dynamics of helix winding captured by isolated single filaments within microdroplets}. 

\begin{figure}
    \centering
    \includegraphics[width=5.2in]{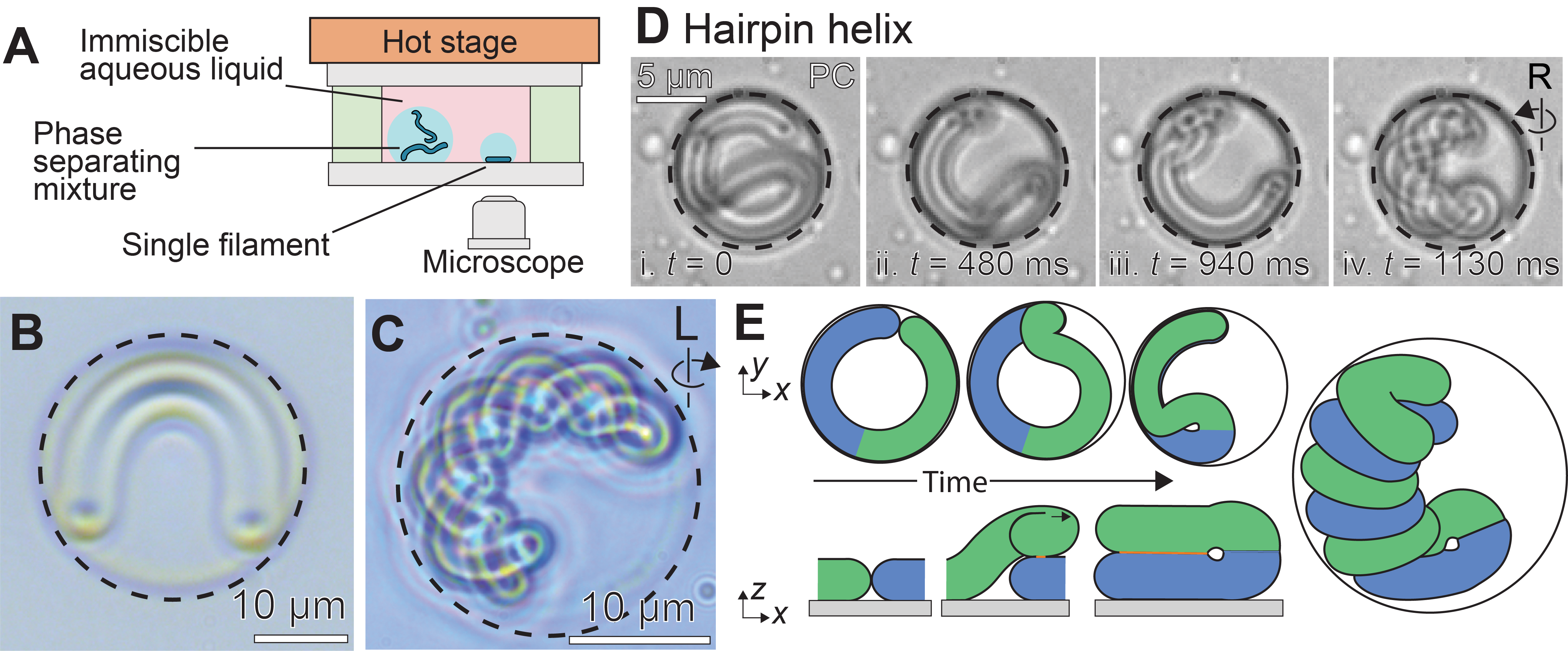}
    \caption{\textbf{Helical winding dynamics of isolated filaments in microdroplets.} \textbf{A} Microdroplets are formed by Rayleigh-Plateau breakup of our binary fluid within an immiscible refractive-index-matched aqueous fluid (Methods). \textbf{B} Sufficiently small microdroplets exhibit the nucleation and growth of single filaments, which grow and evolve in isolation. Larger microdroplets can contain multiple condensate droplets, which sometimes assemble into ``proto-networks'' (Fig. \ref{SIfig:microdroplet}). \textbf{C} Helical coils form in isolation, but only when filaments self-contact. \textbf{D} High speed imaging shows dynamics of coil winding following formation of ribbon contact. \textbf{E} Schematic of ribbon zippering and coil winding process.}
    \label{fig:Microdroplets}
\end{figure}

To investigate the physical mechanism driving this coiling, we confine individual filaments within microdroplets (Methods)---allowing us to directly observe the nucleation and growth of free-floating filaments in isolation with high speed imaging. The same binary mixture is injected into an immiscible aqueous fluid (glycerol and water), resulting in the droplet breakup into microdroplets of radius $R_\mathrm{MD}=3$ to $100~\mathrm{\muup m}$, which pin to the bottom glass substrate to form hemispherical microdroplets (Fig. \ref{fig:Microdroplets}A). Within sufficiently small microdroplets ($R_\mathrm{MD}\lesssim30~\muup \mathrm{m}$), we can observe the nucleation and growth of single filaments, which grow and evolve in isolation (Fig. \ref{fig:Microdroplets}B), and in some cases wind to form free-floating helical coils, similar to those observed in the bulk Hele-Shaw cells (Fig. \ref{fig:Microdroplets}C). We confirm that filament hydrodynamics from filament growth do not appreciably affect filament conformation within these microdroplets (Fig. \ref{SIfig:microdroplet} E and F), allowing us to use this well-controlled environment to directly test the conditions for coiling. 

As before, we do not see any critical temperature beneath which filaments spontaneously coil; filaments exist at all temperatures within the binodal, and for up to an hour when held at a set temperature $58\degree\text{C}$ deep within the binodal ($\Delta T\approx 10\degree\text{C}$). Instead, we only observe coiling events when filaments self-contact, which occurs when the filament lengthens enough to wrap around the circumference of the microdroplet to allow the hemispherical caps to contact. The initiation of coiling solely at the moment of incipient self-contact, rather than below a set critical temperature, demonstrates that coiling is a physical conformational transformation, rather than a mesophase transition.

High speed imaging of filaments during this self-contact demonstrate the dynamics of this coiling (Fig. \ref{SIfig:microdroplet}D, Movies 5 and 6). As with network linking in the Hele-Shaw cells, coiling is preceded by the arrested coalescence of filaments (Fig. \ref{fig:Microdroplets}Di) which zipper into a metastable ribbon (ii). Once zippered, the ribbon winds to initiate the helix formation (iii). After a full turn is wound, the helical region rapidly propagates, forming a double helix with a hairpin turn at one end (iv). We again observe no preference in helix handedness (Fig. \ref{fig:Microdroplets}C and D are opposite handedness), supporting that the initial winding is a stochastic spontaneous symmetry breaking event, leading to chirality at the macroscale---despite remaining in the achiral smectic phase at the microscale. The formation of helical coils within microdroplets, where hydroynamics are absent, further supports that filament adhesion and coiling must be strictly energetically-favorable conformational rearrangements.

\begin{figure}
    \centering
    \includegraphics[width=5.7in]{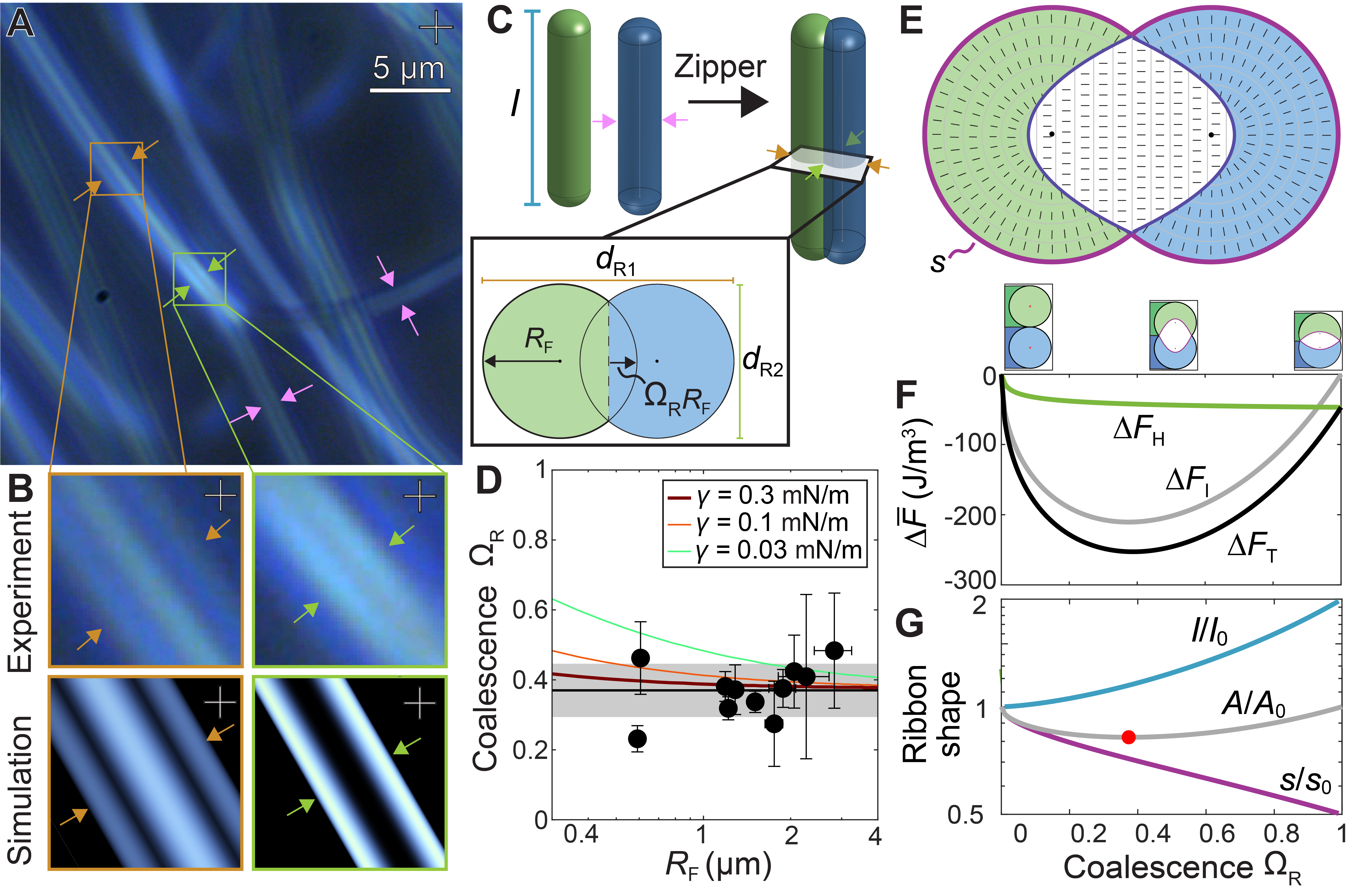}
    \caption{\textbf{Ribbons form by partial coalescence of filaments}. \textbf{A} Birefringence pattern of ribbon (orange arrows) zippering from two free filaments (pink arrows). \textbf{B} Closeup shows two dark axial bands when viewing the major ribbon diameter and a single dark axial band when viewing the minor diameter, in agreement with the simulated birefringence for the proposed ribbon microstructure. \textbf{C} Rendering of free filaments and their arrested coalescence as a ribbon. Filaments colored arbitrarily. Inset shows schematic cross-section with definition of major diameter $d_\text{R1}$ (orange arrows), minor diameter $d_\text{R2}$ (green arrows), and nondimensional coalescence $\Omega_\text{R}$. \textbf{D} Arrested coalescence measured from ribbon $d_\text{R1}$ and free filament radii $R_\text{F}$ for ribbons whose zippering is captured on video. Error bars reflect one standard deviation over five manual measurements. Black line and grey shaded region indicate average and standard deviation $\Omega_\text{R}=0.37\pm0.07$. Lines show theoretical prediction for $\Omega_\text{R}$ for different values of $\gamma$, which converge to be indistinguishable from the black line for $\gamma\gg K/R_\text{F}$. \textbf{E} Proposed smectic microstructure within ribbon. Constant layer spacing requires two parabolic domain walls (purple) separating domains of radially-bent (colored) and straight (uncolored) smectic layers. Cross sectional contour $s$ shown in pink. \textbf{F} Modeled free energy density during coalescence for $K=20~\text{pN}$, $\gamma=0.3~\text{mN/m}$, and $R_\text{F}=1~\muup\text{m}$. Helfrich energy decreases monotonically and cannot arrest coalescence. Interfacial energy exhibits geometrically-determined minimum, which dominates arrested coalescence. Minima of this model plotted for several values of $\gamma$ in panel $\textbf{D}$ (colored lines). Both energies taken in reference to $\Omega_\text{R}=0$, which is that of the uncoalesced straight filaments, $\bar{F}_\text{I,F}=2\gamma / R_\text{F}$ and $\bar{F}_\text{H,F}= \left(K/\left(2R_\text{F}^2\right)\right)\text{ln}\left(R_\text{F}/R_\text{C}\right)$, where the defect core size $R_\text{C}\approx 0.1~\text{nm}$ is estimated from the molecular layer spacing \cite{morimitsu2024spontaneous}. \textbf{G} Geometric-minimum $\Omega_\text{R}^\text{Theory}\approx0.374$ (red dot) set by competition between increasing $l$ and decreasing $s$ during coalescence. Minimum is independent of $R_\text{F}$.}
    \label{fig:ribbon}
\end{figure}

\textbf{Arrested coalescence of filaments stabilizes ribbons}. We measure the transient conformation of filaments during coalescence to quantify interfacial and elastic interactions. Figure \ref{fig:ribbon}A shows one such example of a ribbon in the process of zippering. The free filament ends of diameter $2R_\text{F}$ (pink arrows in lower right)  merge in a Y-junction. Out-of-plane twisting of the ribbon allows measurement of both the major $d_\text{R1}$ and minor $d_\text{R2}$ diameters of the double-barreled structure (indicated by green and orange arrows, respectively, in Fig. \ref{fig:ribbon}A--C). We observe minor diameters $d_\text{R2}\approx2R_F$, and major diameters $d_\text{R1}$ larger than $2R_\text{F}$ and smaller than $4R_\text{F}$ (Fig. \ref{fig:ribbon}B), suggesting that filaments partially coalesce where they contact, forming a double-barreled structure (Schematized in fig. \ref{fig:ribbon}C). We define a nondimensional extent of coalescence $\Omega_\text{R}\equiv \left(4R_\text{F} - d_\text{R1}\right)/\left(2R_\text{F}\right)$, bounded by $0$ and $1$ for filaments at incipient contact and full coalescence, respectively. We measure many filaments which form ribbons and obtain a constant $\Omega_\text{R}\approx0.37 \pm 0.06$, indicating that coalescence of filaments is arrested, stabilizing the ribbon structure (Fig. \ref{fig:ribbon}D). Notably, this $\Omega_\text{R}$ is largely independent of $R_\text{F}$ in the range of observed radii, indicating that filaments have a well-prescribed cross-sectional aspect ratio. 

This arrested coalescence is reminiscent of similar phenomena in polymer gels and yield-stress fluids, where elasticity of an internal elastic network can balance the reduction in interfacial energy \cite{pawar2012arrested,domenech2015rheology,dahiya2016arrested,dahiya2017arrested,kaufman2017photoresponsive}. Coalescence dynamics of nematic droplets are similarly mediated by the elastic stresses associated with rearrangements of the internal mesophase microstructure, which compete with the driving force of surface tension \cite{terentjev1995nematicdroplets,terentjev2005nematicdroplets,oleg1998nematicdroplets}. Motivated by arrested coalescence in these distinct systems, we consider how the interfacial area and elasticity of the smectic microstructure might evolve under the constraints of smectic coalescence to form a similar energetic minimum or kinetically-jammed state. 

Unlike arrested coalescence in isotropic gels and nematic liquid crystals, the smectic liquid crystal must retain constant layer spacing throughout the transient coalescence process. Indeed, these smectic constraints are responsible for the kinetic barriers preventing droplets from reshaping into spheres, and thus should be expected to play salient roles in dictating droplet shape transformations. Within ribbons, the liquid crystal structure must rearrange to accommodate the unusual double-barreled boundary geometry while maintaining both constant layer spacing and homeotropic anchoring at the interface. Homeotropic anchoring requires layers to exhibit a sharp change in orientation close to the contact line between the two barrels of a ribbon, inducing curvature singularity defects in smectic order \cite{LavrentovichKlemanFCDenergetics,Lacaze2025PRL}. In sessile droplets and films, similar curvature singularities self organize into parabolic domain walls, which uniquely allow curved layers to join with flat layers without layer compression \cite{kamien2024sessiledroplets, wei2022focal,Canevari2023Minimizer,Lacaze2005ThinFilm,Lacaze2006ThinFilm,kim2016controlling}.  These domain walls can form a range of low-energy defect structures, including pure-kink points, dislocations, oily streaks, or focal conic flowers \cite{wei2022focal,kamien2024sessiledroplets,Lacaze2005ThinFilm,Lacaze2006ThinFilm,meyer2009focal,beller2013focal} that decorate the wall to accommodate boundary conditions, generally with a negligible energy penalty \cite{kim2016controlling}. 

We propose that similar parabolic domain walls can form running axially down the ribbon (dark purple lines in figure \ref{fig:ribbon}E) separating regions of distinct smectic layer bend. The domain walls separate two external regions where layers bend to match the curvatures of each ribbon barrel (green and blue in figure \ref{fig:ribbon}E) from an internal region of undistorted smectic layers (uncolored). To the best of our knowledge, this is the only geometrically-permissible arrangement consistent with the constraints of homeotropic anchoring \cite{morimitsu2024spontaneous} and constant layer spacing \cite{Lacaze2025PRL} for the observed ribbon interface geometry.

We use the observed optical textures of ribbons to check consistency of our proposed mesophase microstructure. Birefringence of ribbons exhibits two dark axial bands when viewed perpendicular to the major axis (orange in Fig. \ref{fig:ribbon}B) and a single dark band when viewed parallel to the major axis (green). We compare these experimentally-observed textures to simulated textures. We use an established open-source Jones-Matrix calculation tool \cite{chen2024lcpom} to simulate light propagation through our proposed liquid crystal microstructure in quantitative true-color (Methods). We simulate the resulting birefringence patterns at a wide range of viewing angles, for multiple filament radii and $\Omega_\text{R}$ to demonstrate insensitivity of the general features to these changes (Fig.  \ref{fig:filamentsensitivity}--\ref{SIfig:DHrotation}). Representative patterns are shown for comparison in Fig. \ref{fig:ribbon}A, in good agreement with the experimental images: when viewing the major diameter, the simulated ribbon shows two dark axial bands; when viewing along the minor diameter, the ribbon shows a single dark band along the axial centerline. Crucially, this microstructure can evolve continuously during the partial coalescence of two filaments: smectic layers shrink and dissolve only at the tip of the domain wall (Fig. \ref{fig:layerVanish}), circumventing the need to overcome energy barriers for discontinuous layer generation or deletion present in other shape transformations \cite{weinan1999dynamics,arora1989reentrant}. Smectic fluidity within layers allows lengthening along the long axis to conserve volume during coalescence.

To understand why the coalescence of ribbons is arrested, we model the total free energy $F_\text{T}$ of the ribbon for varying $\Omega_\text{F}$ and $R_\text{F}$ using the Helfrich energy $F_\text{H}$ \cite{Helfrich1973} with the addition of interfacial energy $F_\text{I}$:

\begin{align}
    F_\text{T}&= F_\text{I} + F_\text{H}\nonumber\\
    &=\gamma A + \frac{K}{2}\int_\mathscr{V} H^2\text{d}V  .\label{Maineq:energy}
\end{align}

\noindent The first term is the energy of the interface between the homeotropically-anchored smectic phase with interfacial area $A$ and interfacial tension $\gamma\sim0.3~\mathrm{mN/m}$ estimated from Hansen Solubility Parameters \cite{murase2023hansen, browne2025diversity} ({Methods}). The second term is the Helfrich distortion energy associated with bend in the smectic layers of mean curvature $H\left(\textbf{x}\right)$, integrated over the volume of the condensate $\mathscr{V}$, penalized by the bend modulus $K\approx 20~\text{pN}$ \cite{Browne2025Bending}. We assume the domain walls contribution to energy is negligible, as is typical \cite{kim2016controlling}. The imposed shape of the ribbon interface and domain wall fully specifies both $A$ and $H\left(\textbf{x}\right)$, which we numerically compute for varying $\Omega_\text{R}$ (SI \S\ref{sec:ribbonEnergetics}). 

The Helfrich energy decreases monotonically with $\Omega_\text{R}$ (green line in fig. \ref{fig:ribbon}F), indicating that elasticity of the smectic alone cannot be responsible for arresting coalescence. Indeed, curvature energies are generally subdominant to interfacial energies in shaping smectic interfaces \cite{kim2016controlling}. In contrast, the interfacial energy exhibits a radius-independent minimum (grey line in fig. \ref{fig:ribbon}F), which dominates the total energy (black line), which quantitatively matches the $\Omega_\text{R}=0.37$ measured experimentally. The origin of this geometric minimum comes from the constrained evolution of ribbon shape with this microstructure. Our measurements of ribbons suggest that coalescence results in minor ribbon diameter unchanged from the original filament diameter ($d_\text{R,2}=2R_\text{F}$)---consistent with the reported low radial permeation current, which inhibits thickening \cite{arora1989reentrant,weinan1999dynamics}. As the cross-sectional area of the ribbon decreases during coalescence, the ribbon length increases by volume conservation (SI \S\ref{sec:ribbonEnergetics}). We observe this lengthening qualitatively in experiments, where the snapping together of filaments into a ribbon is accompanied by a local buckling, presumably to accommodate the rapid increase in length (Movie 2). The interfacial area $A$ is the product of this increasing filament length (blue line in figure \ref{fig:ribbon}G) and decreasing cross-sectional contour (pink), whose balance gives a purely-geometric minimum (Eq. \ref{eq:RibbonMinimum}), implicitly given by:

\begin{equation}
    0=\pi\left(1-2\Omega_\text{R}\right)+(1-\Omega_\text{R})\sqrt{\Omega_\text{R}(\Omega_\text{R}-2)}+\left(1-2(1-\Omega_\text{R})^2\right)\text{acos}\left(1-\Omega_\text{R}\right)
    \label{Maineq:SurfaceTensionGeom}
\end{equation}

\noindent which gives a theoretical expectation of $\Omega_\text{R}^\text{Theory}\approx0.374$ with no free parameters, in striking agreement with our measurements. This quantitative agreement strongly supports that arrested coalescence of filaments into ribbons is dominated by the geometrically-constrained decrease in interfacial energy. 

Notably, our model predicts a constant $\Omega_\text{R}$ when $R_\text{F}\gg K/\gamma$, with much smaller filaments expected to coalesce more. We do not observe this increase in $\Omega_\text{F}$ at lower $R_\text{F}$ for our observed filament sizes, consistent with the modeled dependence for the estimated $\gamma\sim0.3~\text{mN/m}$ (red line in fig. \ref{fig:ribbon}E). This model also provides a quantitative lower-bound on the interfacial tension $\gamma\gtrsim0.1~\text{mN/m}$, as lower values exhibit a strong $R_\text{F}$-dependence inconsistent with our measurements (colored lines in fig. \ref{fig:ribbon}E). 

\textbf{Ribbon coiling further reduces interfacial area}. Once formed, ribbons quickly wind into coils, generally within $\lesssim1 ~\text{s}$. We hypothesize that coils are similarly driven by arrested coalescence to reduce interfacial area and form regions of undistorted smectic mesophase.

\begin{figure}
    \centering
    \includegraphics[width=5.7in]{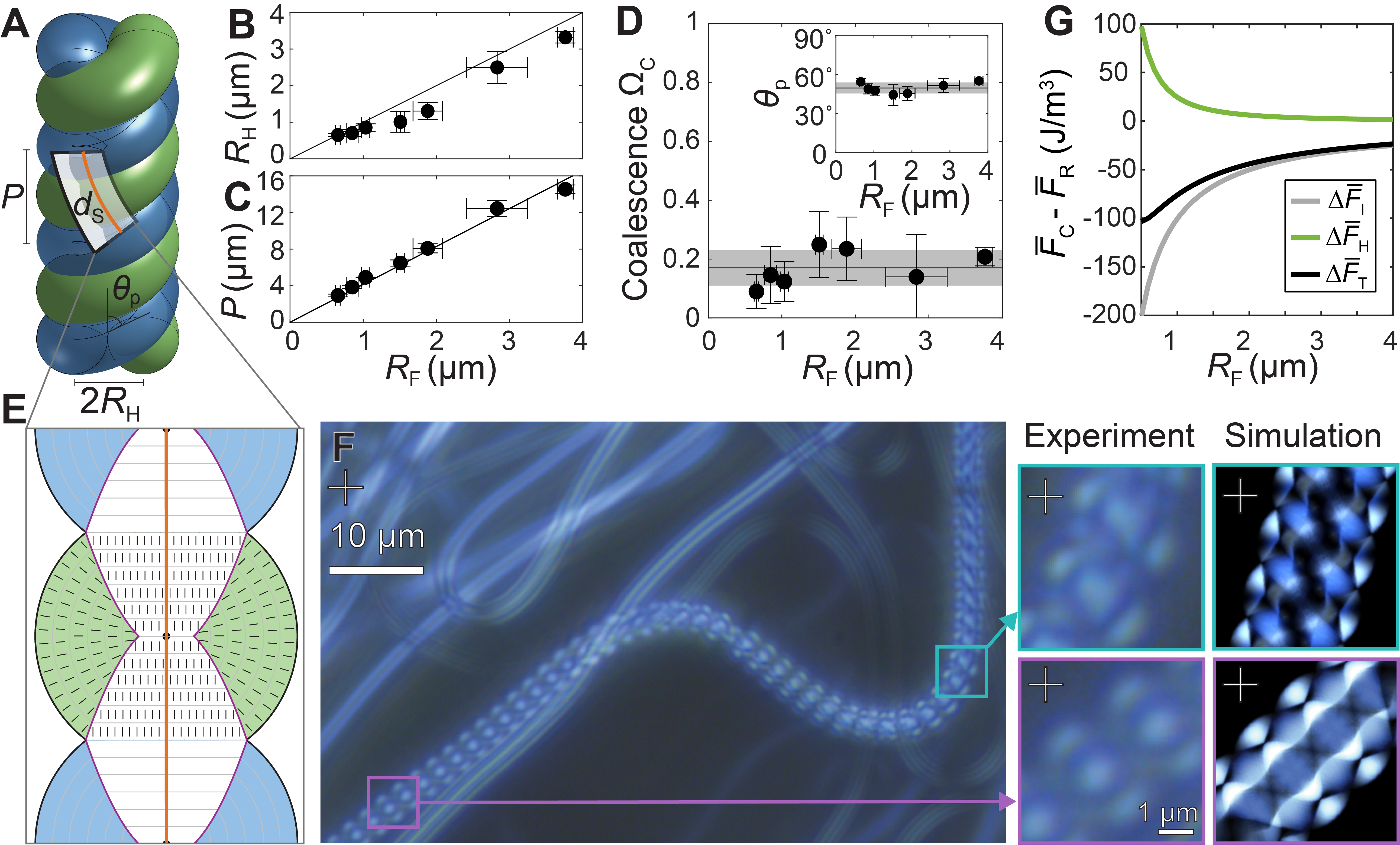}
    \caption{\textbf{Coils increase contact with consistent extent of partial coalescence}. \textbf{A} Rendering of a helical coil with definitions of pitch $P$, pitch angle $\theta_p$, and helix radius $R_\text{H}$. Filaments colored green and blue arbitrarily; blue filament is displayed translucent to make helical backbone visible (black line).  \textbf{B} Helical radius. Black line indicates the condition for maximally-tight coils $R_\text{H}=R_\text{F}$. \textbf{C} Helix pitch $P$ increases linearly with $R_\text{F}$. \textbf{D} Extent of partial coalescence measured from $R_\text{F}$, $R_\text{H}$, and $P$ gives $\Omega_\text{C}=0.17\pm0.06$ \textbf{E} Schematic of proposed internal mesophase structure. Orange line shows contour running between filament centers; our proposed structure treats this contour as a straight chord for simplicity. \textbf{F} Double-helical coil bent by background active flows displaying birefringence at different angles with respect to crossed-polarizers ($0\degree$--$90\degree$). Zoom inset shows segments with helix axis at roughly $45\degree$ and $0\degree$ offset from polarizer for comparison to simulated birefringence for our proposed internal structure. \textbf{G}
Free energy density computed by equation \ref{eq:energy} shows that coils at $\Omega_\text{C}=0.17$ are favorable to ribbons at $\Omega_\text{R}=0.37$ for the range of observed $R_\text{F}$ (SI \S\ref{sec:ribbonEnergetics} and \S\ref{sec:coilEnergetics}). }\label{fig:coil}
\end{figure}

We quantify the arrested coalescence within coils by measuring the helical tube radius $R_\text{H}$ (Fig. \ref{fig:coil}B) and pitch $P$ (Fig. \ref{fig:coil}C). The helical tube radius indicates that coils are maximally-tight, $R_\text{H}\approx R_\text{F}$ (black line in fig. \ref{fig:coil}B), with little space in the coil core. We define the coalescence between neighboring coils $\Omega_\text{C}\equiv 1-d_\text{s}/\left(2R_\text{F}\right)$, where $d_\text{s}$ is the center-to-center distance between neighboring coils along the curved helical tube (orange line in fig. \ref{fig:coil}A). By construction, $\Omega_\text{C}$ is bound by $0$ and $1$ for helical coils with incipient filament contact and full coalescence, respectively.  We compute $\Omega_\text{C}$ from $R_\text{F}$ before coiling and $P$ and $R_\text{H}$ after coiling using geometric relations of helical coils (\S\ref{sec:coilGeom}). We observe a nearly constant $\Omega_\text{C}=0.17\pm0.06$ (Fig.\ref{fig:coil}D), confirming that coils similarly exhibit arrested coalescence with a well-prescribed aspect ratio. The constant coalescence and helical tube diameter predict a constant pitch angle $\theta_\text{p}=\text{acos}\left(2(1-\Omega_\text{C})/\pi\right)\approx54\degree$ to $66\degree$, in reasonable agreement with measured values of $\approx44\degree$ to $55\degree$ in networked coils (Fig.\ref{fig:coil}D inset).

We propose an analogous microstructure composed of parabolic domain walls similar to those in ribbons, which separate regions of bent smectic layers along the coil periphery from a single annular core of undistorted smectic that runs along the helical backbone (Fig. \ref{fig:coil}E). We test the consistency of this microstructure by again simulating the birefringence pattern (Methods) using open-source Jones-Matrix calculation tool \cite{chen2024lcpom}. Our simulated birefringence patterns show stronger sensitivity to viewing angle (Fig. \ref{SIfig:DHrotation}), but nevertheless exhibit reasonable agreement with experimental observations for coils, though their precise azimuthal viewing angle is unknown (Fig. \ref{fig:coil}F). Unlike the microstructure of ribbons, this microstructure in coils does not provide a smooth pathway for $\Omega_\text{C}$ to vary, since there are no vanishingly small layers that can delete barrier free. Instead, we presume that ribbons rearrange into coils while conserving the number of layers, giving a theoretical prediction of $\Omega_\text{C}^\text{Theory}\approx\Omega_\text{R}/2=0.19$, in reasonable agreement with the experimental value of $\Omega_\text{C}=0.17\pm0.06$ (Supplement, Fig. \ref{fig:layerVanish}B).

The spontaneous winding of ribbons into coils suggests a reduction in free energy density. Similar coiling phenomena arise from entropic interactions of highly flexible filaments \cite{snir2005entropically,snir2007helical}, where the geometry of a double helix allows filaments to optimally overlap depleted volume; though the coiling we study here is driven by energetic rather than entropic interactions. Additionally, our proposed microstructure suggests that significant portions of coil interiors remain undistorted, despite the exterior of the coil having significant added helical curvature. We hypothesize that both of these factors contribute to the favorable energetics of coils. To test this hypothesis, we numerically integrate equation \ref{Maineq:energy} in helical coordinates (SI \S\ref{sec:coilEnergetics}) to obtain the free energy density of coils with $\Omega_\text{C}=0.17$, which we compare to the free energy density of ribbons with $\Omega_\text{R}=0.37$ (Fig. \ref{fig:coil}G). At all $R_\text{F}$ the interfacial energy is lower in the coiled state than the ribbon state, as expected from analogies to entropic coiling. In contrast, the Helfrich distortion energy increases during coiling, driven by the added curvature of the helical path for smectic layers near the periphery; though this effect is attenuated by the undistorted interior regions. For the estimated magnitudes of $K\approx20~\text{pN}$ and $\gamma\approx0.3~\text{mN/m}$, the interfacial energy dominates for the full range of observed $R_\text{F}$ (black line), indicating coils are expected to be energetically favored over ribbons. 

Taken together, the corroboration between experiments, modeling, and simulated textures all suggest that filament zippering into ribbons and winding into coils is driven purely by interfacial energy, under the constraints of smectic layering. This physical mechanism provides a route for nascent linkages to form between filaments during the networking of smectic condensates.

% -----------------------------------

\section*{Discussion}
Our results demonstrates that complex coiled morphologies can form in condensates driven purely by interfacial tension and constraints on smectic layering. Demixing smectic liquid crystals are kinetically trapped as filaments \cite{naito1995preferred,naito1997pattern,todorokihara2004periodic} by limitations on radial growth of new smectic layers \cite{weinan1999dynamics,arora1989reentrant,pratibha1992cylindrical}, precluding relaxation into spherical droplets. Here, we observe these filaments can spontaneously decrease their interfacial area instead by partially-coalescing into double-barreled ``ribbons'', and then winding into coils. The preference to remain in these complex morphologies, rather than spheres with minimized interfacial area, suggests the importance of local energetic minima and kinetic trapping by the geometric constraints of the liquid crystal morphology.

We propose one possible set of microstructures that is consistent with this interpretation. The proposed microstructures provide theoretical predictions for the point of arrested coalescence in ribbons and coils, in excellent quantitative agreement with measurements. Simulations of the birefringence patterns through these microstructures also exhibit reasonable agreement with experimental images. These microstructures suggest a mechanism whereby filaments can undergo smooth morphological transformations to reduce interfacial area under the constraints of smectic layering. Our modeling supports that the transformation from filaments to coils decreases interfacial energy, outweighing the increase in elastic distortion. The initial winding of ribbons into coils likely has an energy barrier associated with increasing this bend before the interfacial area can be reduced: phenomenologically, this barrier seems to be sufficiently small, as ribbons typically wind on short $\lesssim1~\text{s}$ timescales. We speculate that this barrier could be overcome by the sudden lengthening of ribbons as they zipper from filaments, which imposes confining stresses, or from imbalances in the lengthening between slightly asymmetric filaments. 

In principle, other lower-energy microstructures may be possible within ribbons and coils; for example by forming more exotic defect or dislocation structures along the domain wall, as seen in other liquid crystals \cite{wei2022focal,kamien2024sessiledroplets,Lacaze2005ThinFilm,Lacaze2006ThinFilm,meyer2009focal,beller2013focal}. Our proposed microstructure provides an upper bound on the energy, and validates that an energetic minimum can be expected. This microstructure further justifies how a smectic can spontanously form a macroscopically-chiral structure, without the driving force of molecular-scale chirality.

This arrested coalescence and coiling has long-term affects on the coarsening of demixing smectic condensates. Dense filaments can align, partially coalesce, and coil to form new linkages, which form new nodes of the growing network. We do not resolve the particular pathway whereby metastable coiled domains further coarsen as the nascent network structure relaxes. Initial compaction of some of these coils shows the formation of supercoils and more complex knotted structures, and eventually flat drops (``bulged discs''), which dominate the late-time network. We anticipate that similar microstructural rearrangements and interfacial area reductions may underlie much of this compaction, but scattering through these irregular 3D structures precludes direct quantification of coarsening. Characterizing the coarsening dynamics of these aggregates remains an important step in quantitatively understanding the final network architecture.

The physical stresses that drive condensate networking may also play some role in other biological or engineered systems, where condensates can form more complex networks. For example, liquid-liquid phase separation of biological proteins in living systems can play important roles in the formation of tight junction networks \cite{pombo2024membrane}, spatiotemporal transcription networks \cite{hnisz2017phase}, and cytoskeletal networks \cite{basu2025vimentin}. Our observations demonstrate one physical mechanism that could contribute to network formation, in addition to active stresses, when the condensing fluid forms a smectic or otherwise lamellar liquid crystal mesophase. Characterizing when these stresses play appreciable roles comparable to active stresses in these more complex living systems remains an open question. 

These results also elucidate potential avenues for material design. The emergent organization of mesoscale networks is set by an interplay of thermally-controlled growth, steric and hydrodynamic controlled confinement, and rearrangements between metastable condensate conformations. Quantitative understanding and control over the nonequilibrium dynamics of this assembly could provide designed control over condensate networks as templates for mesoporous materials.

\section*{Materials and Methods}
\noindent\textbf{Chemicals}. All experiments use a mesogenic liquid crystal 4'-cyano 4-dodecyloxybiphenyl (\textit{TCI Chemicals}), commonly referred to as 12OCB, which in isolation is known to form a smectic A liquid crystal mesophase directly upon cooling from the isotropic liquid. We mix this with a solvent squalane (\textit{TCI Chemicals}), which is a branched nonpolar alkane. All mixtures are prepared at high temperature $>90\degree$C, where all fluids are homogeneous well-mixed isotropic liquids. 

\noindent\textbf{Imaging}. We image phase separation using an inverted light microscope (\textit{Ziess}) with a $5\times$, $20\times$, or $63\times$ magnification objective, as denoted in figure captions. We image with brightfield (BF), phase contrast (PC), and crossed-polarizers (indicated by cross on images). Unless otherwise noted, all images are recorded at $1$ to $30$ frames per second on a $2448\times2048$ color CMOS (\textit{Thorlabs} Kiralux). High speed imaging is instead captured at 1000 frames per second on a $1280\times960$ monochrome high speed CMOS (\textit{Phantom} VEO 1010).

\noindent\textbf{Microdroplet Confinement}.  We prepare microdroplets containing binary smectogen (12OCB) solvent (squalane) mixtures by injecting a heated stream of the binary liquid mixture through a fine needle (25G) into a heated aqueous bath of 89 wt.\% Glycerol 11 wt.\% Water. The 12OCB-squalane mixture is completely immiscible with the aqueous bath, resulting in breakup of the stream into microdroplets of varying radii $R_\text{MD}\sim1$ to $100~\mathrm{\muup m}$, which rise by buoyancy to the top air-aqueous fluid interface. A second cover slip is placed on top and held in place by UV-cured glue. Cover slip spacing is maintained by $63~\mathrm{\muup m}$ diameter silica particles suspended in the UV-glue. Microdroplets make contact with this second cover slip, and remain in contact after the entire Hele-Shaw cell is inverted for imaging. We confirm that microdroplets remain in contact with the second (bottom) cover slip after inversion by repeating this procedure with rhodamine 6G added to the aqueous phase and imaging with confocal microscopy (\textit{Leica} Stellaris).

\noindent\textbf{Simulated optical textures of helical coils.} We simulate the optical response of partially coalesced single and double helical coils modeled using the proposed internal smectic microstructure. The simulated images are compared to experimental birefringence under crossed polarizers. Helices generated with different filament radii and pitch angles are studied to demonstrate the sensitivity of the optical signature to these variables. 

We numerically impose the helical backbone of each coil, using parameters extracted from experimentally observed structures. We discretize the helical path into $1000$ points per pitch length. At each discretized point, we define a plane normal to the helical tangent vector, approximating the plane drawn in Fig. \ref{fig:coil}E. In each plane, we prescribe a director field on a local rectilinear grid according to the proposed partially-coalesced microstructure in Fig.\ref{fig:coil}F.  For numerical ease, we simplify the smectic microstructure by projecting the flat 2D microstructure schematized in figure \ref{fig:coil}E along a helical path; though in reality, the green contour should be curved along a perpendicular axis along the helical tube, introducing some errors into our simulated birefringence texture. This defined director field is then linearly interpolated onto a $400\times400\times400$ rectilinear grid to ensure an even density of directors appropriate for simulation. Despite the error of this microstructural simplification, our simulated birefringence patterns show reasonable agreement with experimental observations for coils (Fig. \ref{fig:coil}F). 

We simulate the optical birefringence texture under polarized optical microscopy using the open-source \textit{LCPOM} simulation package \cite{chen2024lcpom}. This software package simulates light propagation through the anisotropic director field to reconstruct optical textures as they appear under crossed polarizers, using known properties of 4'-cyano 4-pentylbiphenyl (5CB), a similar liquid crystal. Refractive index differences with 12OCB may contribute to differences in simulated and experimental textures. Images are calculated in full spectrum color. We simulate textures for filaments (Fig. \ref{fig:filamentsensitivity}), ribbons (Fig. \ref{fig:ribbonsensitivity}), and double-helical coils (Figs. \ref{fig:SI-sim-sensitivity-double} and \ref{SIfig:DHrotation}), all for varying geometric parameters and polarizer-analyzer angles. 

\noindent\textbf{Estimation of interfacial tension.} 

We estimate the interfacial tension between 12OCB and squalane using Hansen Solubility Parameters (HSP) \cite{browne2025diversity} and the empirical correlation proposed by Murase \textit{et al.} \cite{murase2023hansen}. The smectic mesophase is known to exhibit two distinct surface tensions for homeotropic $\gamma_\perp$ and planar $\gamma_\parallel$, respectively. We estimate $\gamma_\parallel\approx3~\text{mN/m}$ for the HSP differences between squalane and the whole 12OCB molecule. We estimate $\gamma_\perp\approx0.3~\text{mN/m}$ using the HSP difference between squalane and an effective dodecanol tail region of 12OCB, which represents the region of the molecule that the solvent interacts with during hometropic anchoring, following our previous methods validated by X-ray scattering \cite{browne2025diversity}. The large anisotropy in these interfacial tensions reflects the strong homeotropic anchoring condition, making the relevant interfacial tension $\gamma=\gamma_\perp\approx0.3~\text{mN/m}$. This falls well-within the expected general bounds of $\gamma\gg0.01~\text{mN/m}$, estimated for single-component smectic-isotropic interfaces at phase equilibrium \cite{pratibha1992cylindrical}, and $\gamma\ll6~\text{mN/m}$, measured for smectic-immiscible liquid interfaces \cite{harth2013measurement}. Because of this wide range and lack of direct experimental measure for $\gamma$ for our particular smectic-solvent interface, we test several values of $\gamma$ to show its effect on the model (colored lines in Fig. \ref{fig:ribbon}D). Our experimental observation of a constant $\Omega_\text{R}\approx0.37$ for $R_\text{F}>0.5~\muup\text{m}$ suggests a lower-bound interfacial tension of $\gamma\gtrsim0.1~\text{mN/m}$, consistent with our theoretical estimate.

\noindent\textbf{Data, Materials, and Software Availability}. All study data are included in the article and/or supporting information. Code used to generate simulated optical birefringence textures is available at the online repository \href{https://github.com/Yzagzag/Smectic-Helices}{https://github.com/Yzagzag/Smectic-Helices}.

\noindent\textbf{Acknowledgments}. 
It is our pleasure to thank Randall D. Kamien for insightful discussions, and Paulo E. Arratia for access to the confocal microscope. The authors acknowledge NSF support through DMR-2309043. This material is also based in part on work supported by the NSF Graduate Research Fellowship Program (to P.G.S.) under Grant No. DGE-1845298. C.A.B. acknowledges postdoctoral fellowship support from the Center for Soft and Living Matter at the University of Pennsylvania.

\noindent\textbf{Author contributions.} {C.A.B. and C.O.O. designed the experiments; C.A.B., A.B., and Y.C. performed all experiments and analyzed all data; C.A.B., P.G.S., and J.C. designed the theoretical models; Y.Z. developed new code; C.A.B., P.G.S., A.G.Y., and C.O.O., discussed the results and implications, and wrote the manuscript; C.A.B. and C.O.O. designed and supervised the overall project.}

\noindent\textbf{Author ORCIDs.}
Christopher A. Browne 0000-0002-3945-9906;
Paul G. Severino 0000-0002-1648-4148;
Yvonne Zagzag 0000-0001-7600-7821;
Jacob Z. Cloutier 0009-0000-6465-0633;
Aaron C. Boyd 0009-0005-7179-0921;
Yihao Chen 0000-0002-9956-5719;
Arjun G. Yodh 0000-0003-4744-2706;
Chinedum O. Osuji 0000-0003-0261-3065;\\

\noindent \textbf{Movie Captions}

\noindent \textbf{Movie 1} Collapse and linkage of inter-tangled filaments to form new network aggregate node. Video captured on color CMOS under crossed polarizers (0$\degree$ and 90$\degree$). System is cooled at a constant rate $0.3\degree\text{C/min}$ from $T=69\degree\text{C}$. Video begins at $T\approx62.3\degree\text{C}$ viewed at 5$\times$ magnification ($1413\muup\text{m}\times1413\muup\text{m}$ field of view), then switches to black when switching to 20$\times$ magnification ($353\muup\text{m}\times353\muup\text{m}$ field of view) to show close up of collapse event, then back to 5$\times$ magnification at end to show large-scale modifications to network. Playback $3\times$ real time.

\noindent \textbf{Movie 2} Collapse and linkage of inter-tangled filaments to form new network aggregate node. Video captured on monochrome high speed CMOS using phase contrast imaging at 63$\times$ magnification ($101\muup\text{m}\times76\muup\text{m}$ field of view). Playback real time.

\noindent \textbf{Movie 3} Free floating single helical coil composed of a single filament wound with itself. Video tracks center of coil as it rotates and advects in background flow. Captured at 63$\times$ magnification on color CMOS under crossed polarizers (0$\degree$ and 90$\degree$). $5~\muup\text{m}$ scale bar visible, playback real time.

\noindent \textbf{Movie 4} Free floating double helical coil composed of a single filament doubled back to wind with itself. Video tracks center of coil as it rotates and advects in background flow. Captured at 63$\times$ magnification on color CMOS under crossed polarizers (0$\degree$ and 90$\degree$). $5~\muup\text{m}$ scale bar visible, playback $2\times$ real time.

\noindent \textbf{Movie 5} Filament ribbon confined to microdroplet winding to form helical coil. Video captured on monochrome high speed CMOS using phase contrast imaging at 63$\times$ magnification ($40\muup\text{m}\times40\muup\text{m}$ field of view). Playback $1/6\times$ real time.
 
\noindent \textbf{Movie 6} Another example of filament ribbon confined to microdroplet winding to form helical coil. Video captured on monochrome high speed CMOS using phase contrast imaging at 63$\times$ magnification ($48\muup\text{m}\times48\muup\text{m}$ field of view). Playback $1/3\times$ real time.

\noindent \textbf{Movie 7} Filament ribbon confined to microdroplet winding anti-aligned to form knotted pretzel-like structure. Video captured on monochrome high speed CMOS using phase contrast imaging at 63$\times$ magnification ($48\muup \text{m}\times48\muup \text{m}$ field of view). Playback real time.

% -----------------------------------
\bibliography{references}
\bibliographystyle{unsrt}

\newpage

\renewcommand\thefigure{S\arabic{figure}}
\renewcommand\theequation{S\arabic{equation}}
\renewcommand\thesection{S\arabic{section}}
\renewcommand\thesubsection{S\arabic{subsection}}
\noindent\textbf{\Large Supplementary Information: \\Arrested coalescence drives helical coiling and networking of filamentous smectic condensates}
\setcounter{secnumdepth}{4}
\setcounter{page}{1}
\setcounter{figure}{0}
\setcounter{section}{0}
\newline

\section{Extended figures}
\begin{figure}[hbt!]
    \centering
    \includegraphics[width=5.3in]{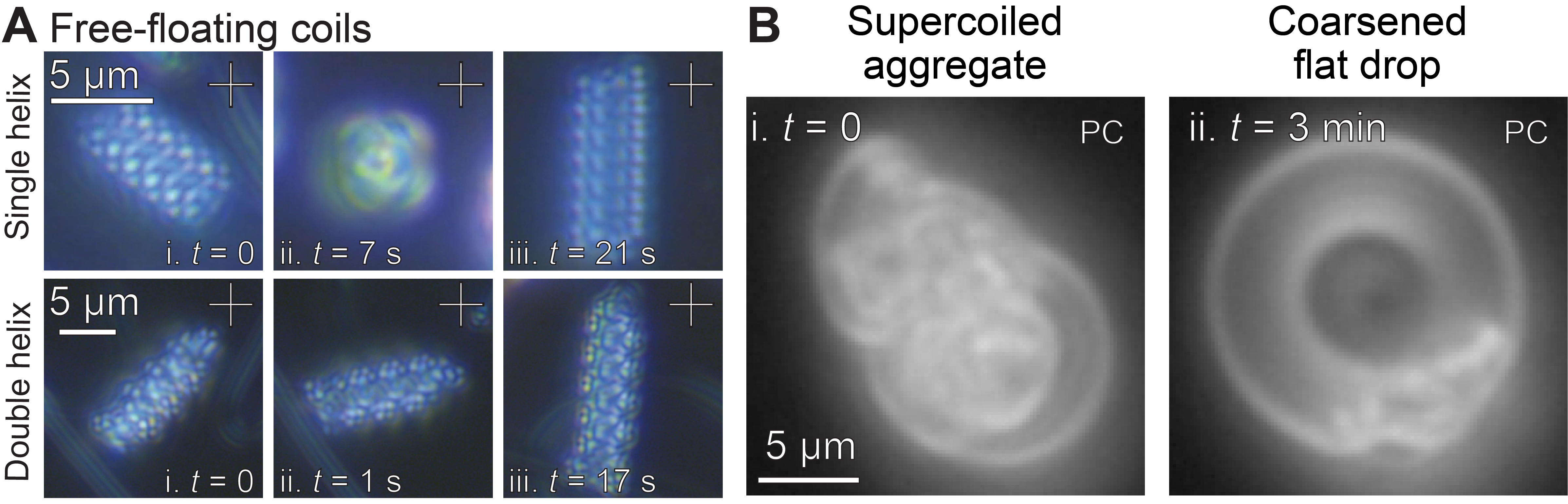}
    \caption{\textbf{A} Free-floating helical coils, comprised of single filaments wound into a single helix (top; Movie 3) or into a double helix (bottom; Movie 4). These free-floating helices are relatively rare and do not play any obvious role in the network formation, but conveniently provide multiple viewing angles of the coils as they rotate and grow while advected by the background flows of network activity. \textbf{B} These free-floating coils can persist for tens of minutes, until they stochastically rearrange to form free-floating supercoiled aggregates, which coarsen over the course of minutes. These free-floating supercoiled aggregates are analogous to the aggregate nodes in the larger network.}
    \label{SIfig:freefloat}
\end{figure}

\begin{figure}[b]
    \centering
    \includegraphics[width=5in]{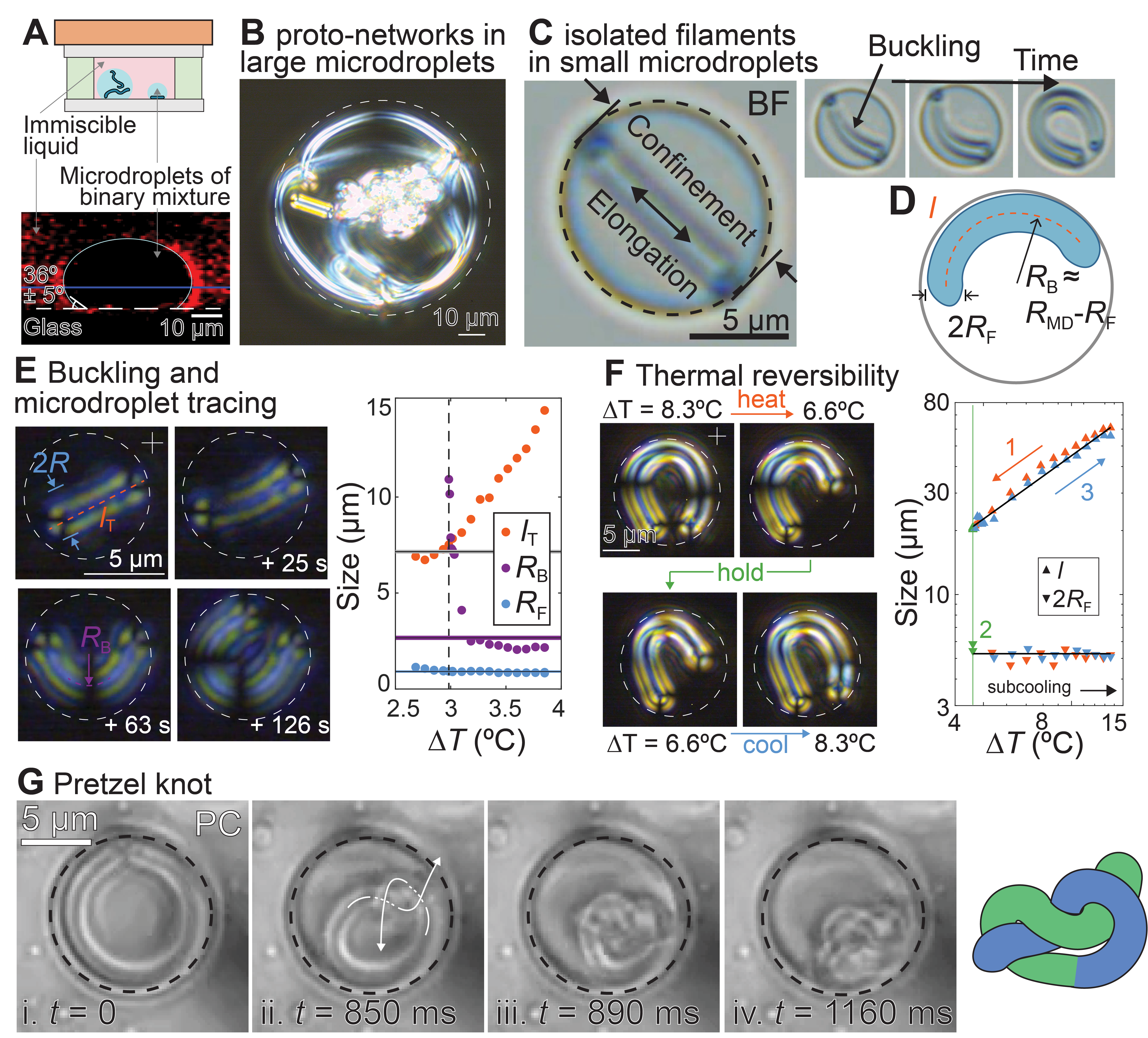}
    \caption{\textbf{A} Schematic of microdroplets in Hele-Shaw cell. Confocal z-slice in aqueous fluid dyed with rhodamine-6G demonstrates microdroplets are pinned to the glass substrate. \textbf{B} Sufficiently large microdroplets exhibit multiple condensate nucleations, which can assemble into ``proto-network'' structures similar to those observed in Hele-Shaw cells. \textbf{C} Sufficiently small microdroplets exhibit nucleation of a single filament, which lengthens and bends when it comes into contact with the microdroplet interface, after which it buckles, with no measurable deformations of the microdroplet interface. \textbf{D} Schematic of filament dimensions in microdroplet, used in plots in panels E and F. \textbf{E} As the filament length exceeds the radius of the microdroplet, filaments bend and follow the curvature of the microdroplet until the curvature of the filament matches that expected for the confining microdroplet size (horizontal purple line). Buckling in response to microdroplet confinement indicates that hydrodynamics from growth are negligible in microdroplets. \textbf{F} Filament configurations are quasi-statically reversible: alternating cycles of heating and cooling at $0.5\degree\mathrm{C/min}$  lead to quasi-static lengthening and shortening of filaments at constant radius. This reversibility indicates filament volume, configuration, and aspect ratio are all in quasi-static equilibrium. \textbf{G} In addition to double helices, we also observe the formation of knotted structures, which can form when the two free ends of a filament contact and partially-coalesce while anti-aligned (Movie 7), demonstrating the generic preference to form varied coiled motifs to increase the adhesive contact area. }
    \label{SIfig:microdroplet}
\end{figure}

\clearpage

\section{Geometric relations of partially coalesced helical coils}\label{sec:coilGeom}

For any helical path, the pitch $P$, pitch angle $\theta_p$, and tube radius $R_\text{H}$ can be related by unwinding the helix contour:

\begin{equation}
    \text{tan}\left(\theta_p\right)=\frac{2\pi R_\text{H}}{P}~~~~~\text{(any helix)},
\end{equation}

\noindent We compute the expected pitch $p$ of helically coiled filaments of radius $R_\text{F}$ partially coalesced by an extent of overlap $\Omega_\text{C}\equiv 1 - d_\text{s}/\left(2R_\text{F}\right)$. The center-center distance between neighboring coils $d_\text{s}$ is measured normal to the helix contour along the curved surface of the helix barrel (orange line in figure \ref{fig:coil}B). The center-to-center distance and helical pitch are similarly related by unwinding the helix contour:

\begin{equation}
    \text{sin}\left(\theta_p\right)=\frac{nd_s}{P}.
\end{equation}
\noindent where $n=2$ is the number of filaments coiled together. Combining these expressions allow the pitch and coalescence to be directly related:
\begin{equation}
    P = \frac{4R_\text{F}\left(1-\Omega_\text{C}\right)}{\sqrt{1-\left(\frac{2R_\text{F}\left(1-\Omega_\text{C}\right)}{\pi R_\text{H}}\right)^2}},
\end{equation}
\noindent or, equivalently:
\begin{equation}
    \Omega_\text{C}=1-\frac{P}{4R_\text{F}}\left(1+\left(\frac{P}{2\pi R_\text{H}}\right)^2\right)^{-1/2}.\label{SIeq:omegaCCompute}
\end{equation}

\begin{figure}
    \centering
    \includegraphics[width=6.5in]{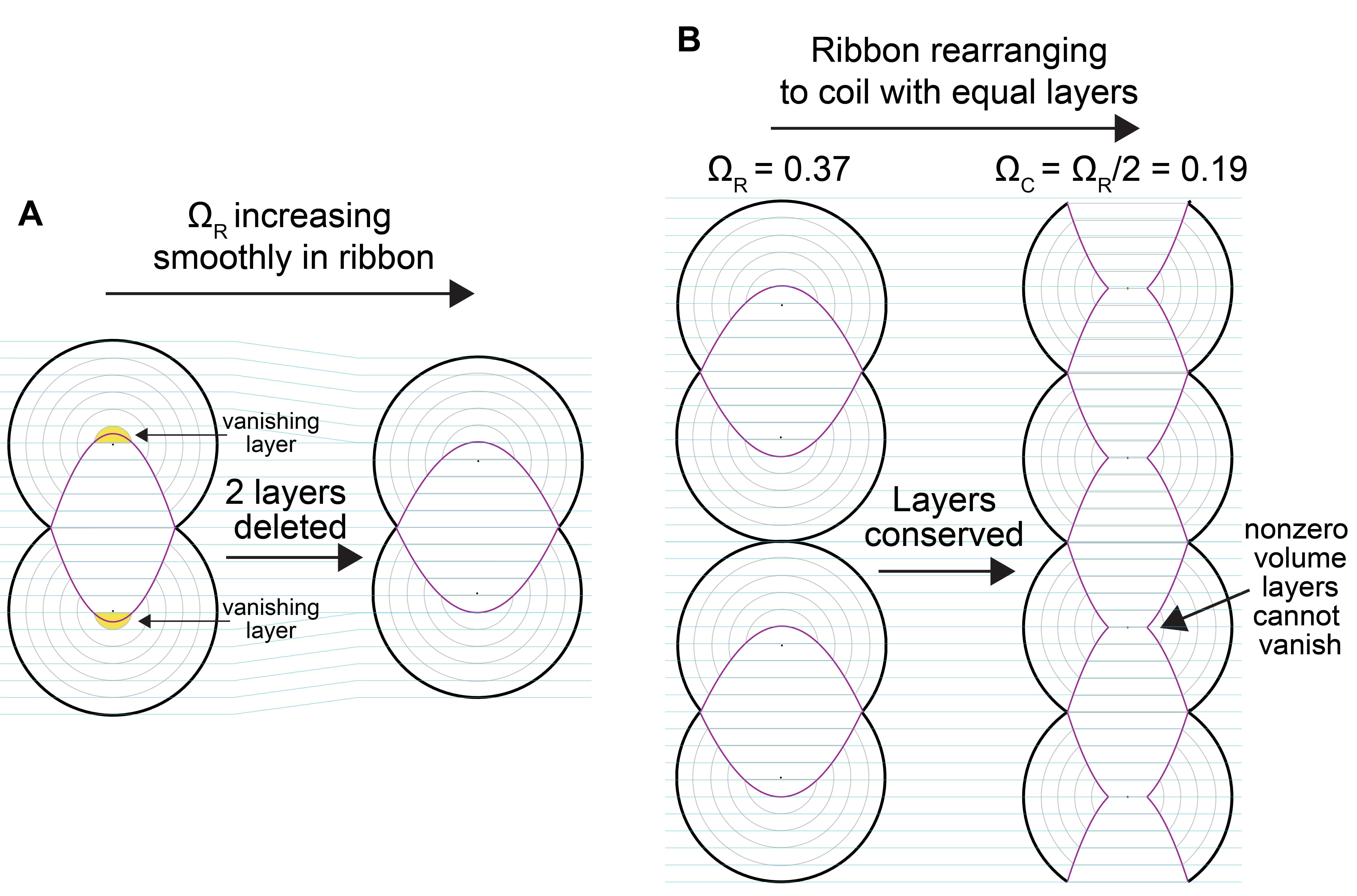}
    \caption{\textbf{A} As filaments coalesce into a ribbon (increasing $\Omega_\text{R}$), smectic layers must delete. The proposed smectic microstructure exhibits layers of vanishing volume at the tips of the parabolic domain walls, which can delete to allow $\Omega_\text{R}$ to increase smoothly from $0$. \textbf{B} Rearrangement of a ribbon to a coil with preserved layers results in $\Omega_\text{C}=\Omega_\text{R}/2$. Once formed, coils do not have vanishingly-small layers that can delete, kinetically trapping them in this configuration.}
    \label{fig:layerVanish}
\end{figure}

\clearpage

\section{Simulated optical textures}
\begin{figure}[h]
     \centering
     \includegraphics[width=6.5in]{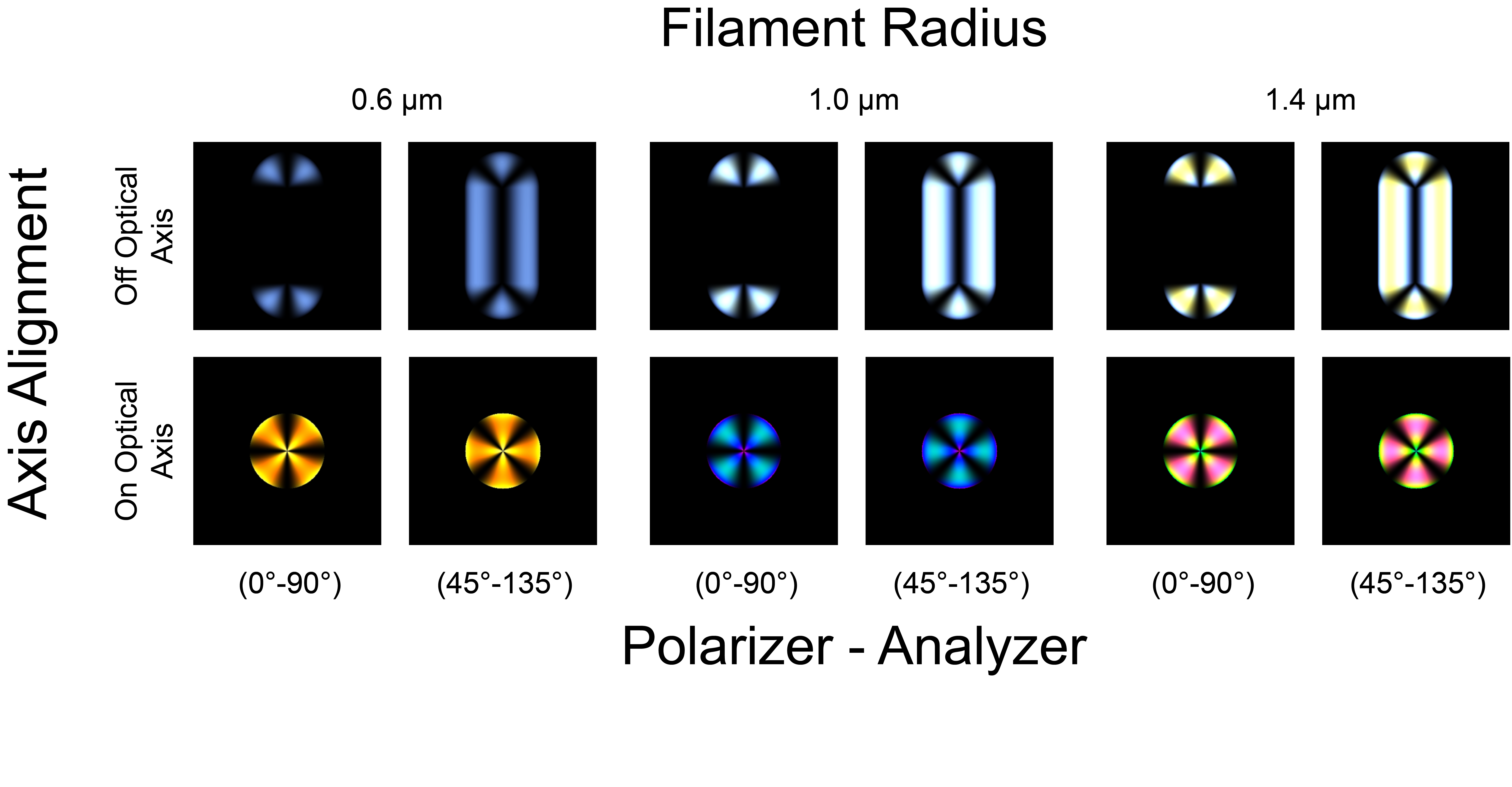}
     \caption{Simulated birefringence patterns through straight filaments for filaments parallel (top row) and perpendicular (bottom row) to optical axis. We test varying polarizer-analyzer angles for filaments of different $R_\text{F}$ to demonstrate sensitivity of texture and color. Good agreement with experimental images of filaments helps validate our implementation of open source code \cite{chen2024lcpom}.}
     \label{fig:filamentsensitivity}
 \end{figure}

\begin{figure}
     \centering
     \includegraphics[width=6.5in]{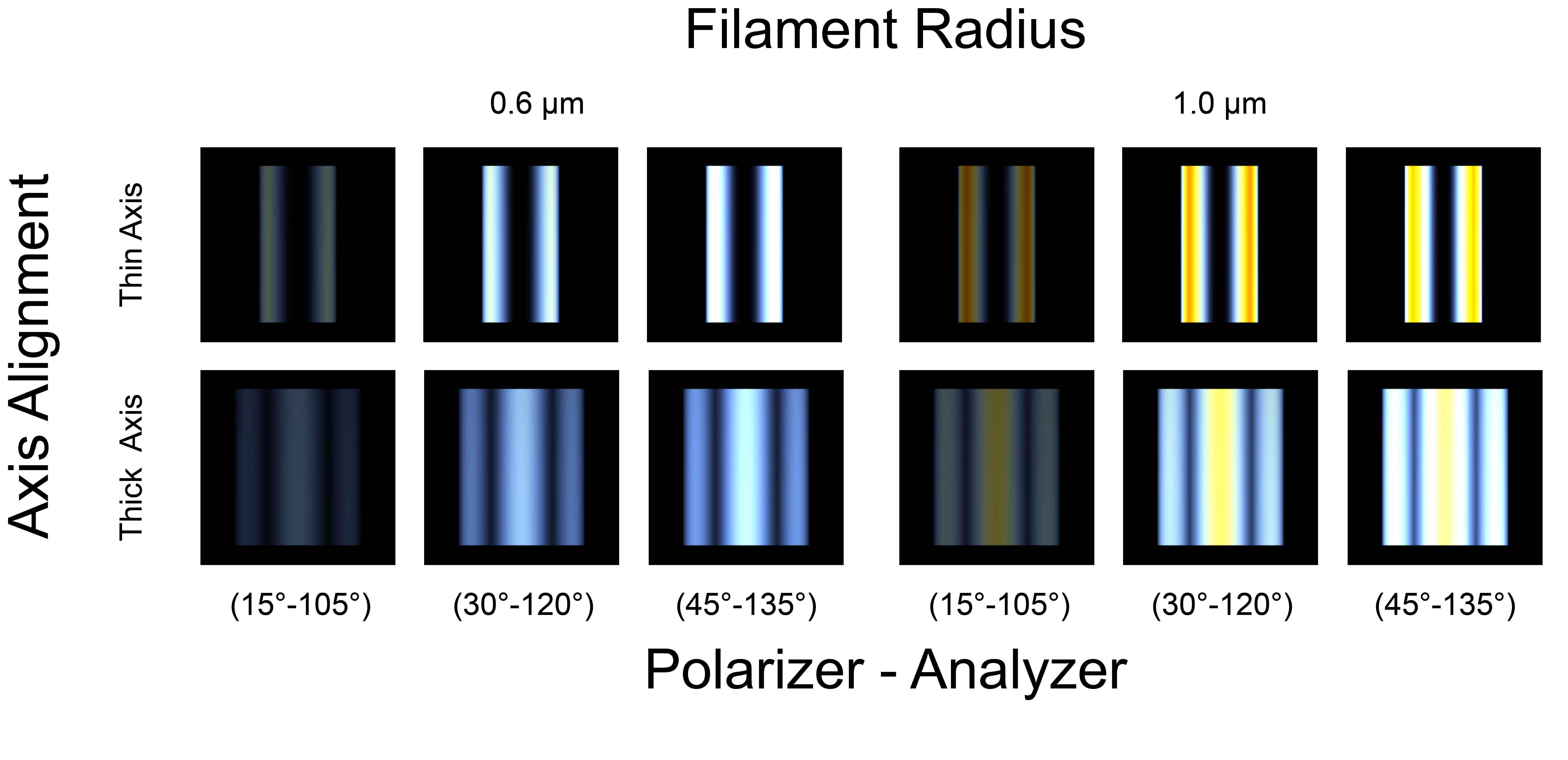} 
     \caption{Simulated birefringence patterns through partially-coalesced ribbons viewed along thin minor axis (top row) and thick major axis (bottom row). We test varying polarizer-analyzer angles for ribbons of different $R_\text{F}$ to demonstrate sensitivity of texture and color. }
     \label{fig:ribbonsensitivity}
 \end{figure}

\begin{figure}
    \centering
    \includegraphics[width=6.5in]{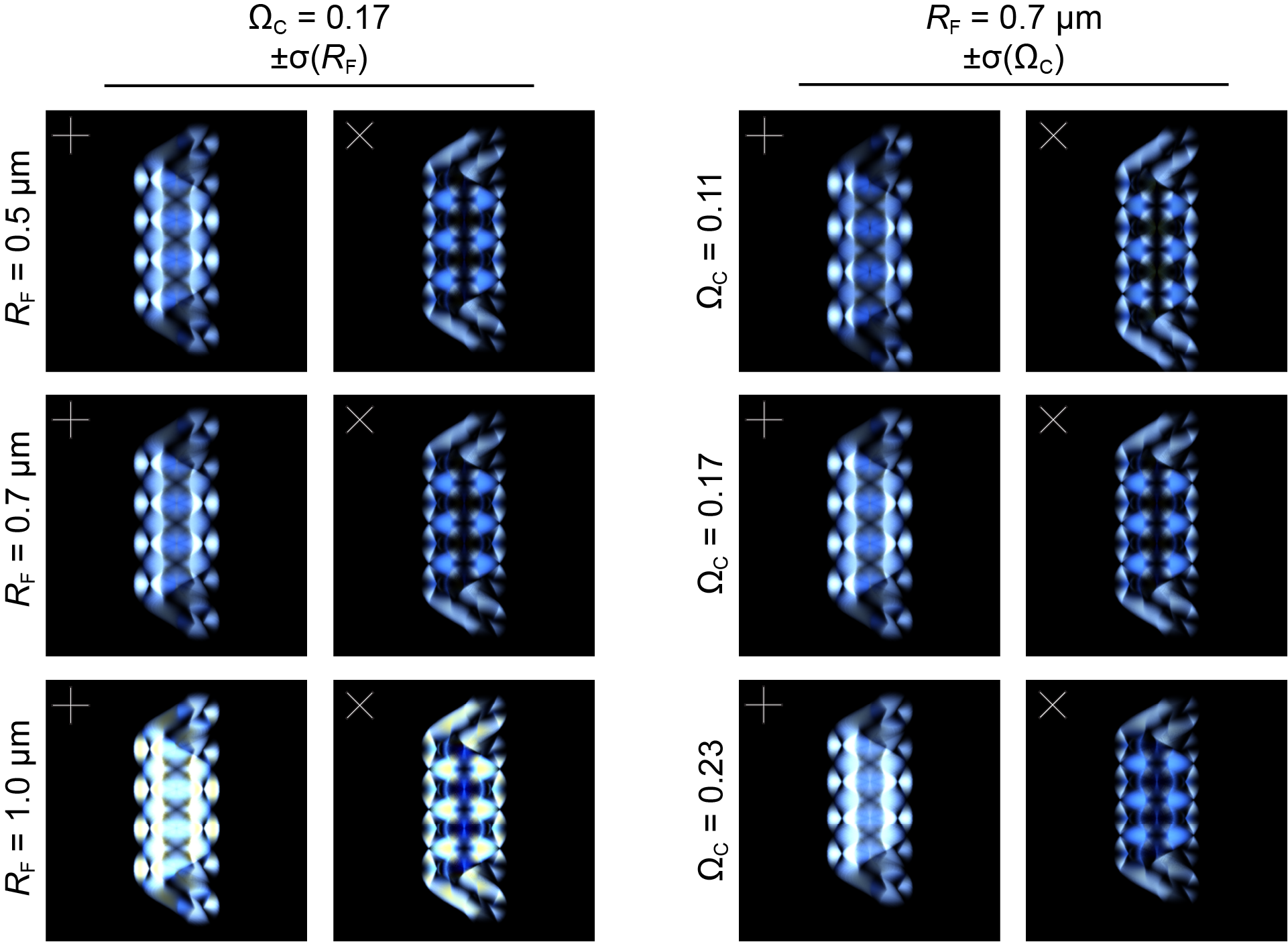}
    \caption{Sensitivity of simulated birefringence patterns through double-helical coils of $R_\text{H}=R_\text{F}$ for plus and minus one standard deviation in filament radius $R_\text{F}$ (from example coil in figure \ref{fig:coil}) (left); and plus and minus one standard deviation in degree of partial coalescence $\Omega_\text{C}$ (right). Within these experimental error bounds, simulated optical textures vary predominantly in intensity and color, with minor variations to qualitative structure.}
    \label{fig:SI-sim-sensitivity-double}
\end{figure}

\begin{figure}
    \centering
    \includegraphics[width=6.5in]{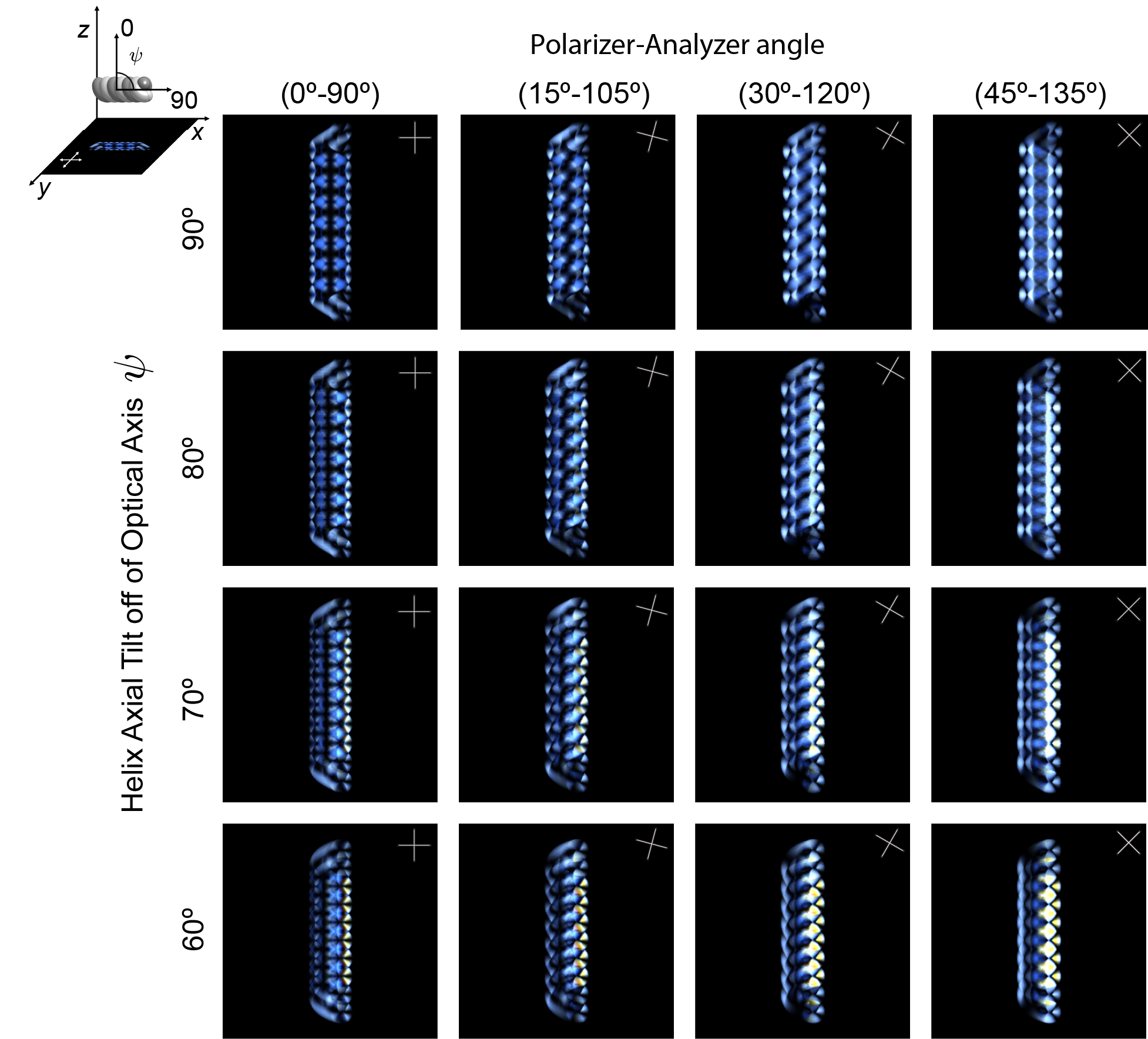}
    \caption{Simulated birefringence patterns through double-helical coil of $R_\text{F}=0.7~\muup\mathrm{m}$ and $R_\text{H}=R_\text{F}$ for varying polarizer-analyzer angles and azimuthal tilt of helical axis. Coil texture is highly sensitive to polarizer-analyzer angle and azimuthal tilt. Insets in figure \ref{fig:coil} are chosen at angles where sensitivity in textures is less pronounced.}
    \label{SIfig:DHrotation}
\end{figure}

For comparison of optical textures, we simulate the light propagation through our proposed mesostructures (Methods) for straight filaments, ribbons composed of partially coalesced filaments, single helical coils, and double helical coils under a wide range of orientations with respect to the optical axis and the polarizer-analyzer. For the straight filaments (Fig. \ref{fig:filamentsensitivity}) and ribbons (Fig. \ref{fig:ribbonsensitivity}), we repeat simulations for slight variations in filament radius ($R_\text{F}$) in order to demonstrate sensitivity to experimental optical measurement error. For double helical coils (Fig. \ref{fig:SI-sim-sensitivity-double}) we additionally test sensitivity to both filament radius ($R_\text{F}$) and helical pitch angle ($\theta_\text{p}$). We additionally simulate the textures of double helices under varying azimuthal angles (out of the imaging plane), for varying angles of  the crossed-polarizers (Fig. \ref{SIfig:DHrotation}). 

\clearpage

\section{Modeled energetics of ribbon during arrested coalescence}\label{sec:ribbonEnergetics}
The arrested coalescence of ribbons and helical coils implies a metastable local minimum in the free energy of filaments during coalescence. To understand this metastable state, we compute the free energy of ribbons for arbitrary extents of coalescence $\Omega_\text{R}\in(0,1)$ using the Helfrich energy:

\begin{align}
    F_\text{T}&= F_\text{I}+F_\text{H}\nonumber\\
    &=\gamma A + \frac{K}{2}\int_\mathscr{V} H^2\text{d}V . \label{eq:energy}
\end{align}

\noindent The terms represent the interfacial energy between the smectic condensate and the surrounding isotropic liquid, and the curvature energy of the smectic mesophase. We assess the change in free energy $\Delta F$ with respect to two straight filaments at incipient contact, whose energetics have been previously modeled \cite{morimitsu2024spontaneous}. As an incompressible fluid, the smectic filaments have a constant density, and hence the total volume of condensate is conserved. We assume these transformations occur at constant radius---supported by observation (Fig. 3A) and the low radial permeation current \cite{arora1989reentrant,weinan1999dynamics}. We therefore compute the varying length of the ribbon, which grows during coalescence, to preserve this constant volume:

\begin{equation}\label{eq:lengthen}
    V=Sl=S_0l_0,
\end{equation}

\noindent where the cross-sectional area of the two filaments is initially $S_0=2\pi R_F^2$, and the cross-sectional area of the single ribbon is:

\begin{equation}
    S_\text{R}=  2R_F^2\left(\pi-\text{acos}\left(1-\Omega_\text{R}\right)+\left(1-\Omega_\text{R}\right)\sqrt{1-\left(1-\Omega_\text{R}\right)^2}\right),
\end{equation}

\noindent\textbf{Helfrich energy in ribbons.} The second term in Eq. \ref{eq:energy} represents the energy associated with distortions of the smectic mesophase. We compute this energy using the Helfrich elastic distortion energy \cite{Helfrich1973} for a smectic A from the average curvature $H$ of the smectic layers, which have a bend modulus $K\approx20~\mathrm{pN}$ \cite{Browne2025Bending}. 

\begin{equation}
    F_\text{H}=\frac{K}{2}\int H^2\text{d}V.
\end{equation}

\noindent For straight filaments, neglecting hemispherical end caps, the average curvature is $H=1/(2r)$ everywhere in cylindrical coordinates. For a single straight filament of length $l_0$ this gives:

\begin{equation}
    F_\text{H,F} =\frac{\pi}{4} Kl_0~\text{ln}\left(\frac{R_\text{F}}{R_\text{C}}\right),
\end{equation}

\noindent where the cutoff length $R_\text{C}\approx0.1~\text{nm}\lesssim d_L=4.8~\mathrm{nm}$\cite{morimitsu2024spontaneous} is the radius of the defect core running axially down the filament, whose defect energy we neglect.

\begin{figure}
    \centering
    \includegraphics[width=0.5\linewidth]{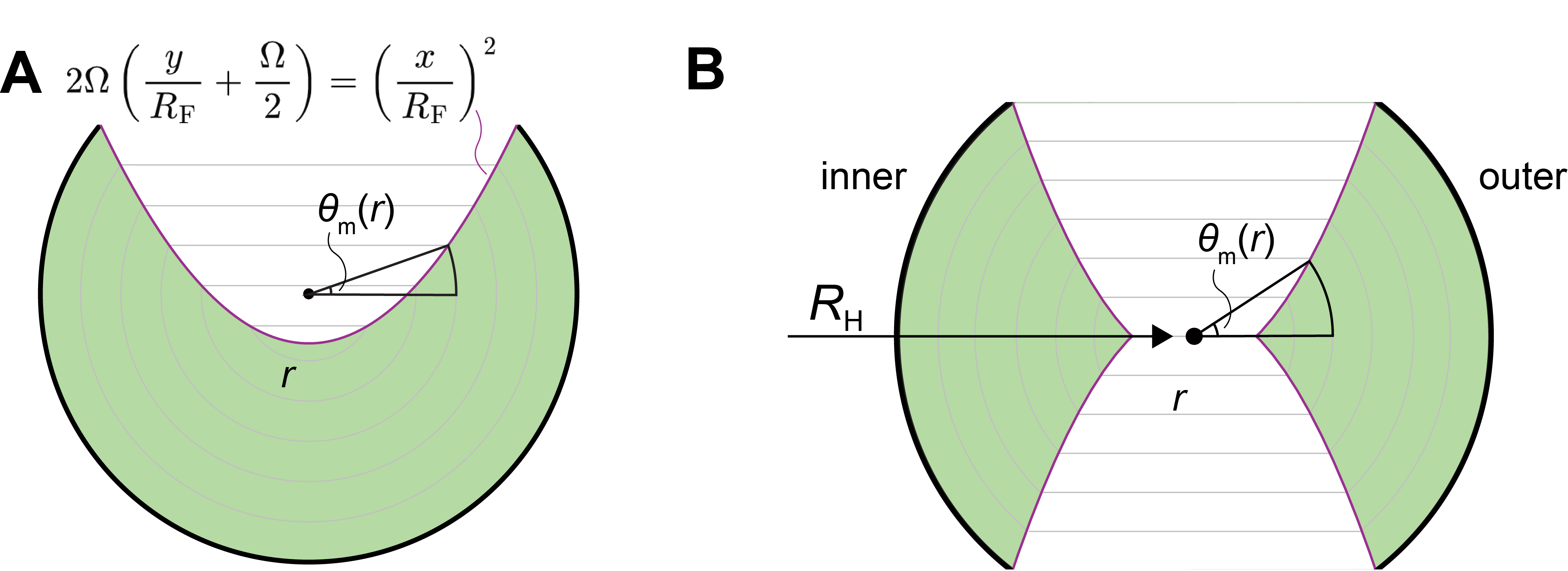}
    \caption{Geometric relations of parabolic domain wall and definition of coordinates for ribbon (\textbf{A}) and coil (\textbf{B}).}
    \label{fig:parabolaGeometry}
\end{figure}

For partially-coalesced ribbons, the same $H=1/(2r)$ exists everywhere outside of the parabolic domain wall, and $H=0$ within the parabolic domain wall. The curvature energy of the ribbon is given by the integration over the area external to the parabolic domain wall (green region in Fig. \ref{fig:parabolaGeometry}A):

\begin{align}
    F_\text{H,~R}&=l\frac{K}{8}\int_{r=\Omega_\text{R}R_\text{F}/2}^{r=R_\text{F}}\int_{\theta=0}^{\theta_m(r;~\Omega_\text{R})}\frac{1}{r}\text{d}\theta\text{d}r\nonumber\\
    &=l\frac{K}{8}\int_{\tilde{r}=\Omega_\text{R}/2}^{\tilde{r}=1}\int_{\theta=0}^{\theta_m(\tilde{r};~\Omega_\text{R})}\frac{1}{\tilde{r}}\text{d}\theta\text{d}\tilde{r},
\end{align}

\noindent where we define $\tilde{r}=r/R_\text{F}$, and $\theta_\text{m}(\tilde{r})$ is the angle where the layers meet the parabolic domain wall in the ribbon,  which is given geometrically by equating circles of radius $\tilde{r}$ with the formula for the parabolic domain wall (Fig. \ref{fig:parabolaGeometry}A):

\begin{align}
    \theta_\text{m}(\tilde{r};\Omega_\text{R}) &=\text{asin}\left(1-\frac{\Omega_\text{R}}{\tilde{r}}\right)
\end{align}

\noindent We integrate this expression numerically to obtain the results in figure \ref{fig:ribbon}E. 

\noindent\textbf{Interfacial energy in ribbons.} The first term in Eq. \ref{eq:energy} represents the interfacial energy. The interfacial area change to partially coalesce two filaments into a ribbon is given by: 

\begin{equation}
    \Delta A_\text{R}= 4R_Fl\left(\pi-\text{acos}\left(1-\Omega_\text{R}\right)\right)-4\pi R_F l_0.
\end{equation}

In the limit where surface tension dominates $\gamma\gg K/R_\text{F}$, the metastable minimum in $\Omega_\text{R}$ is set purely-geometrically by the competition between coalescence, which reduces cross-sectional area, and lengthening, which increases the interfacial area as set by volume conservation under constant $R_\text{F}$. Thus, the metastable local minimum partial coalescence is set by:

\begin{equation}
    0 = \left[\frac{\partial }{\partial \Omega_\text{R}}\left(\left(1+\frac{\left(1-\Omega_\text{R}\right)\sqrt{1-\left(1-\Omega_\text{R}\right)^2}}{\pi-\text{acos}\left(1-\Omega_\text{R}\right)}\right)^{-1}\right)\right]_{\Omega_\text{R}^\text{Theory}},\label{eq:RibbonMinimum}
\end{equation}

\noindent where the functions for $\Delta A\left(\Omega_\text{R}\right)$ and $l\left(\Omega_\text{R}\right)$ have been explicitly inserted. This relationship, which has no free parameters, gives the exact implicit formula:

\begin{equation}
    0=\pi\left(1-2\Omega_\text{R}\right)+\left(1-\Omega_\text{R}\right)\sqrt{\Omega_\text{R}\left(\Omega_\text{R}+2\right)}+\left(1-2\left(1-\Omega_\text{R}\right)^2\right)\text{acos}\left(1-\Omega_\text{R}\right)~~~~~~~~\left(\mathrm{for~~\Omega_\text{R}\neq0~and~\Omega_\text{R}\neq1}\right)
    \label{SIeq:SurfaceTensionGeom}
\end{equation}

\noindent which gives $\Omega_\text{R}^\text{Theory}\approx0.374...$ as the single approximate numerical zero .

\section{Modeled energetics of coil}\label{sec:coilEnergetics}
To compute the interfacial and Helfrich energy in $n-$helical coils, we use the helical coordinate frame radial distance $r\in[0,R_\text{F}]$ within the coil body's circular cross-section, angle $\theta\in[0,2\pi)$ within the coil body's circular cross-section, $\phi\in[0,2\pi]$ parameterizing the angle of one full helical turn. The helical backbone is then parametrized:

\begin{equation}
\textbf{x}_\text{backbone}=\left(x,y,z\right)=\left(R_\text{H}\text{cos}\left(\phi\right),R_\text{H}\text{sin}\left(\phi\right),\frac{P\phi}{2\pi}\right)
\end{equation}
\noindent\textbf{Interfacial energy of helical coil.}
We parametrize the surface of the helical coil by:

\begin{align}
    \mathbf{r}(\phi, \theta, R_\text{F}) = 
    R_\text{H}\begin{bmatrix}
        \text{cos}(\phi) \\
        \text{sin}(\phi) \\
        \frac{P}{R_\text{H}}\frac{\phi}{2\pi} 
    \end{bmatrix}
     + R_\text{F}\text{cos}(\theta)
     \begin{bmatrix}
         \text{cos}(\phi) \\
         \text{sin}(\phi) \\
         0
     \end{bmatrix}
     + R_\text{F}\text{sin}(\theta)
     \begin{bmatrix}
         \text{sin}(\theta_p)\text{sin}(\phi) \\
         -\text{sin}(\theta_p)\text{cos}(\phi) \\
         \text{cos}(\theta_p)
     \end{bmatrix} \label{eq: position def}
\end{align}

The interfacial area of one full rotation (one pitch) of an $n$-helical coil is given by:
\begin{align}
    A &=n\int\int\text{d}A\nonumber\\
    &=2n\int_{\phi=0}^{2\pi}\left(\int_{\theta =0}^{\theta=\theta_o}D_\text{A}\mathrm{d}\theta+\int_{\theta = \theta_i}^{\theta=\pi}D_\text{A}\mathrm{d}\theta\right)\text{d}\phi
\end{align}
\noindent where the symmetric integral over $\theta\in(0,\pi)$ and $\theta\in(\pi,2\pi)$ give a factor of $2$, and the integral over $\phi\in(0,2\pi)$ gives an additional prefactor $2\pi$, and the bounds $\theta_o$ and $\theta_i$ indicating angles where coils contact each other are solved for below. The area element $\text{d}A=D_\text{A}\text{d}\theta\text{d}\phi$ in this coordinate frame is given by the cross product:

\begin{align}
D_\text{A}&=\left|\frac{\partial\textbf{x}}{\partial\theta}\times\frac{\partial\textbf{x}}{\partial\phi}\right|\nonumber\\
&=\frac{R_\text{F}^2R_\text{H}\text{cos}\left(\theta\right)}{\sqrt{\left(\frac{P}{2\pi}\right)^2+R_\text{H}^2}}+R_\text{F}\sqrt{\left(\frac{P}{2\pi}\right)^2+R_\text{H}^2}
\end{align}

\noindent The integration bounds are set by the outermost and innermost angle of contact between neighboring coils: $\theta_o$ and $\theta_i$, respectively, such that $0<\theta_o<\theta_i<\pi$. These angles are given implicitly by an intermediate angle $\psi$:

\begin{align}
    0 = 
      \left(R_\text{H}\pm\sqrt{R_\text{F}^2-\left(\frac{P}{2\pi}\right)^2\left(\frac{\pi/n-\psi}{R_\text{H}}\right)^2\left(\left(\frac{P}{2\pi}\right)^2+R_\text{H}^2\right)}\right)\text{sin}\left(\psi\right)-\left(\frac{P}{2\pi}\right)^2\left(\frac{\pi/n-\psi}{R_\text{H}}\right)\text{cos}\left(\psi\right),\label{SIeq:psi}
\end{align}

\noindent which has exactly 2 real zeros in $\psi\in(0, \pi/2)$, denoted $\psi_\text{zero,1}$ and $\psi_\text{zero,2}$. From these the two contact angles  are given by:

\begin{align}
    &\theta =  \begin{cases} 
      \text{arcsin}\left(\frac{P\left(\pi/n-\psi_\text{zero,i}\right)}{2\pi R_\text{F}\text{sin}\left(\theta_p\right)}\right) & \text{postive~branch}\\%0 < \theta \leq \pi/2 
      \pi - \text{arcsin}\left(\frac{P\left(\pi/n-\psi_\text{zero,i}\right)}{2\pi R_\text{F}\text{sin}\left(\theta_p\right)}\right) & \text{negative~branch}%\pi/2 < \theta < \pi
    \end{cases}
\end{align}

\noindent $\psi_\text{zero,2}$ is always on the negative branch, giving $\pi/2<\theta_i<\pi$. $\psi_\text{zero,1}$ can be on either branch, giving $0<\theta_o<\theta_i$.

The volume of one turn of an $n$-helical coil is given by:

\begin{align}
    V&=n\int\int\int\text{d}V\nonumber\\
    &=\left(4\pi n\right)\int_{\theta=0}^{\theta=\pi}\int_{r=0}^{r=r_m\left(\alpha\right)} D_\text{V}\mathrm{d}r\mathrm{d}\theta
\end{align}

\noindent where the integral over $\mathrm{d}\phi$ gives a prefactor $2\pi$, and the additional factor of $2$ accounts from the equal area contribution on the bounds $\theta\in(\pi,2\pi)$. The volume element $\text{d}V=D_\text{V}\text{d}r\text{d}\theta\text{d}\phi$ in this coordinate frame is given by:

\begin{align}
D_\text{V}=&\left|\frac{\partial\textbf{x}}{\partial r}\cdot\left(\frac{\partial\textbf{x}}{\partial\theta}\times\frac{\partial\textbf{x}}{\partial\phi}\right)\right|\nonumber\\
=&r\left(\left(\frac{P}{2\pi}\right)^2+R_\text{H}^2\right)^{1/2} \nonumber\\
&+\frac{r^2R_\text{H}\text{cos}\left(\theta\right)}{\left(\left(\frac{P}{2\pi}\right)^2+R_\text{H}^2\right)^{1/2}}
\end{align}

\noindent and the radial integration bound is given by:
\begin{equation}
    r_m = 
    \begin{cases} 
          R_\text{F} & \theta<\theta_o\\
          R_\text{F}\frac{\text{cos}\left(\frac{\left(\theta_i-\theta_o\right)}{2}\right)}{\text{cos}\left(\left|\theta-\frac{\left(\theta_o+\theta_i\right)}{2}\right|\right)} & \theta_o\leq \theta < \theta_i\\
          R_\text{F} & \theta_i\leq\theta\\
    \end{cases}
\end{equation}

We then numerically integrate these expressions to obtain the interfacial energy density of a coil at $\Omega_\text{C}=0.17$, which can be taken in reference to a straight filament and a ribbon at $\Omega_\text{R}=0.37\approx2\Omega_\text{C}$ of the same volume $V$:

\begin{equation}
    \bar{F}_\text{I}=\frac{\gamma A}{V} 
\end{equation}

\noindent\textbf{Helfrich energy in coils}
The Helfrich energy in coils is similarly computed from the average curvature of layers in the coil periphery, with no bend in the central annular region, in the above prescribed helical coordinate frame. The average curvature $H=(1/2)(H_1+H_2)$ is the average of both the radial splay contribution $H_1=1/r$, and the curvature induced by the helical coil $H_2$. The latter is given by the curvature of a helix with the same pitch $p$ as the coil's helical backbone, and a helix radius given by $R_\text{H}'\left(r,\theta\right)=R_\text{H}+r\text{cos}\left(\theta\right)$. Using the formula for the curvature of a helix and substituting in the relations for $R_\text{H}$ gives:

\begin{align}
    H_2&=\left(R_\text{H}'\left(r,\theta\right)+\frac{p^2}{2\pi R_\text{H}'\left(r,\theta\right)}\right)^{-1}\nonumber\\
    &=\frac{1}{R_\text{F}}\left(  \frac{2}{\pi\text{cos}\left(\theta_p\right)}\left(1-\tilde{R}_\text{O,~C}\right)+2\pi\text{cot}\left(\theta_p\right)+\frac{r}{R_\text{F}}\text{cos}\left(\theta\right)  \right)^{-1},
\end{align}

\noindent We can then similarly integrate this averaged curvature over one turn of the $n$-helical coil (green region in Fig. \ref{fig:parabolaGeometry}B):

\begin{equation}
    F_\text{H,C}=4\pi n\frac{K}{2}\int_{r=\Omega_\text{C}R_\text{F}}^{r=R_\text{F}}\left(\int_{\theta=0}^{\theta_\text{C}(r)}H^2~D_\text{V}\text{d}\theta+\int_{\theta=\pi-\theta_\text{C}(r)}^{\pi}H^2~D_\text{V}\text{d}\theta\right)\text{d}r,
\end{equation}

\noindent The angle where the layers meet the parabolic domain wall should be computed iteratively from an intermediate angle $\psi$ analogously to Eq. \ref{SIeq:psi}, but for simplicity we instead use the approximation, which is good for $\Omega_\text{C}\gtrsim0.1$:

\begin{equation}
    \theta_\text{C}(r)\approx\text{asin}\left(1-\frac{\Omega_\text{C}R_\text{F}}{r}\right).
\end{equation}

\noindent We then numerically integrate these expressions to obtain the Helfrich energy density of a coil at $\Omega_\text{C}=0.17$, which can be taken in reference to a straight filament and a ribbon at $\Omega_\text{R}=0.37\approx2\Omega_\text{C}$ of the same volume $V$.

\end{document}